\documentclass[10pt,twocolumn,a4paper]{article}

 \usepackage[final,pagenumbers]{cvww}

\usepackage{upgreek}
\usepackage{bm}

\usepackage[pagebackref,breaklinks,colorlinks,allcolors=cvwwblue]{hyperref}
\usepackage{placeins}
\usepackage{multirow}
\usepackage{colortbl}
\definecolor{tab_color}{HTML}{ff4e47}

\def\paperID{35} 

\title{Pi-GS: Sparse-View Gaussian Splatting with Dense $\bm{\pi}^3$ Initialization}

\author{Manuel Hofer \qquad Markus Steinberger \qquad Thomas Köhler\\
Graz University of Technology\\
Austria
}

\begin{document}

\twocolumn[{
\maketitle
\begin{center}
    \includegraphics[width=\linewidth]{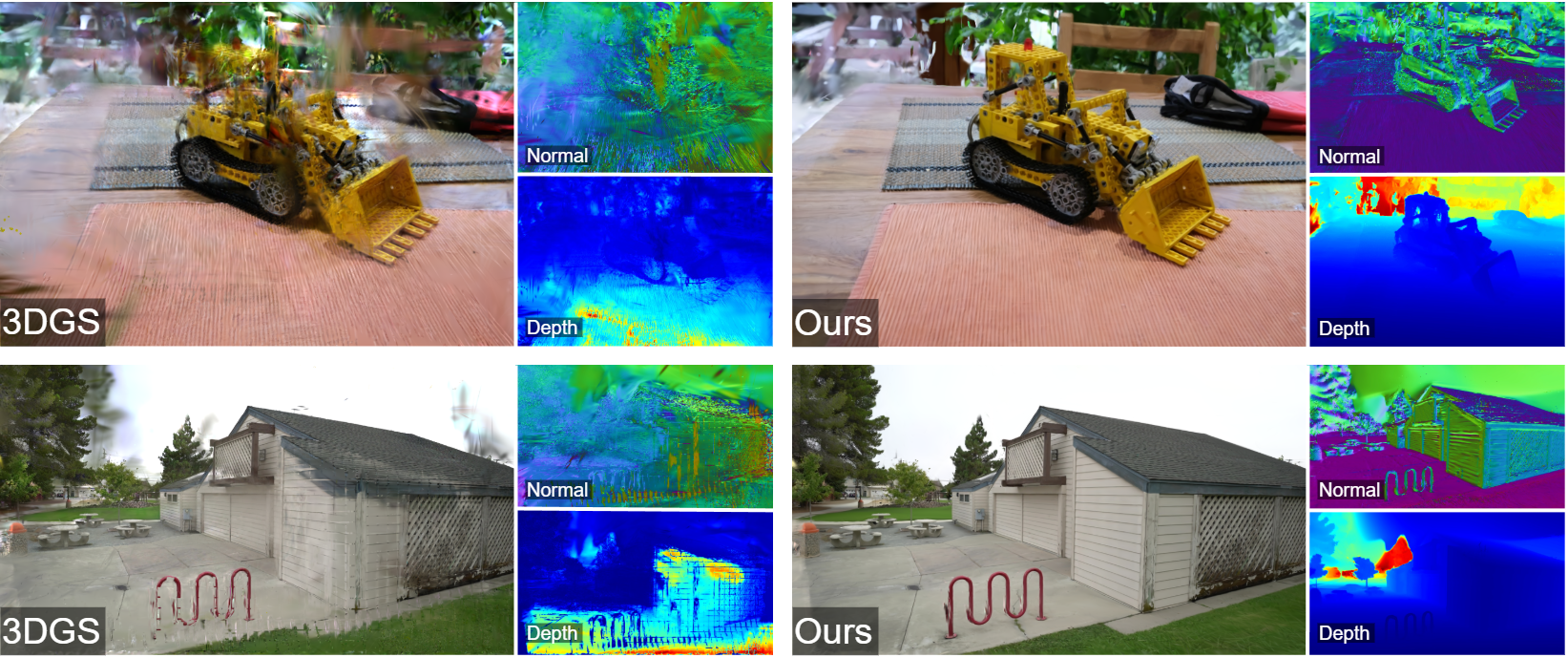}
    \captionof{figure}{3DGS exhibits floaters and view inconsistencies under sparse-view constraints. These artifacts are mostly caused by depth ambiguities and poor Gaussian alignment with the underlying geometry, as shown in the depth and normal maps. By incorporating depth supervision, normal supervision, and additional pseudo views, our method significantly reduces these artifacts and produces more view-consistent novel views with improved Gaussian alignment under sparse-view constraints.}
    \label{fig:placeholder}
\end{center}
}]


\begin{abstract}
Novel view synthesis has evolved rapidly, advancing from Neural Radiance Fields to 3D Gaussian Splatting (3DGS), which offers real-time rendering and rapid training without compromising visual fidelity. However, 3DGS relies heavily on accurate camera poses and high-quality point cloud initialization, which are difficult to obtain in sparse-view scenarios. While traditional Structure from Motion (SfM) pipelines often fail in these settings, existing learning-based point estimation alternatives typically require reliable reference views and remain sensitive to pose or depth errors. In this work, we propose a robust method utilizing $\pi^3$, a reference-free point cloud estimation network. We integrate dense initialization from $\pi^3$ with a regularization scheme designed to mitigate geometric inaccuracies. Specifically, we employ uncertainty-guided depth supervision, normal consistency loss, and depth warping. Experimental results demonstrate that our approach achieves state-of-the-art performance on the Tanks and Temples, LLFF, DTU, and MipNeRF360 datasets.

\end{abstract}
\section{Introduction}
3D scene reconstruction and novel view synthesis (NVS) are rapidly advancing, with many applications across different domains~\cite{EVALScenReconstruction2024}.
These methods can be applied in fields such as Virtual Reality (VR) for creating immersive worlds, cinematography to create visually appealing assets efficiently, or
robot vision to help robots understand their physical environment~\cite{SLAMROBOT2021}.
The foundation of 3D scene reconstruction was laid by traditional Structure from Motion pipelines. More recently, significant advances in NVS were achieved by representing the scene as Neural Radiance Fields (NeRF)~\cite{NeRFOriginal}. These methods achieve state-of-the-art results but suffer from slow training speeds and are unsuitable for real-time rendering due to high latency. Newer methods such as 3D Gaussian Splatting (3DGS)~\cite{3dgs2023} enable high-quality NVS even for real-time rendering. Additionally, training speed is significantly reduced.

A major limitation of these novel view synthesis methods is the need for dense views, which often is not feasible for real-world applications. In sparse-view settings, these methods tend to struggle with bad initialization, depth ambiguities and overfitting to training views. To improve the performance in these settings and counteract the depth ambiguities, certain priors are introduced to better generalize and escape minima throughout the optimization process. Methods such as \textit{DNGaussian}~\cite{DNGaussian} and \textit{Few-shot Novel View Synthesis using Depth}~\cite{few-shotNVS3DGS} leverage monocular depth estimators to regularize the model with the help of the inferred depth. The depth regularization helps significantly to improve the depth ambiguities and increase the generalization capability of the models. A challenge for these models is correct depth scaling, proper point initialization, and accurate camera poses. The initial points and camera poses are traditionally generated using Structure from Motion (SfM) pipelines. However, these pipelines often struggle with sparse input views and limited overlap between views. Recent advancements for sparse-view settings were achieved by leveraging dense initialization with the help of point cloud estimation networks~\cite{InstantSplat, InternGS, SparseGS}. They replace the traditional SfM pipeline with models such as \textit{MASt3R}~\cite{mastrsfm2025} or \textit{DUSt3R}~\cite{dust3r_cvpr24} for the point cloud estimation and camera pose estimation.
The resulting models achieve high-fidelity results but require good initial reference views for accurate predictions. In addition, a time-consuming iterative camera alignment process is required, which can take several minutes. Inaccurate camera poses may further reduce reconstruction quality.

We make the following contributions:
\begin{itemize}
    \item We discuss a method for leveraging a Permutation-Equivariant point cloud estimation network for dense initialization without relying on traditional SfM.
    \item We introduce confidence aware pearson depth loss, to counteract uncertain depth estimations.
    \item We explore the use of PGSR in sparse-view settings for improved geometry alignment and reduced overfitting.
\end{itemize}
Our method achieves state-of-the-art results in sparse-view settings and significantly improves Gaussian surface alignment, while reducing floaters.
Our code is publicly available at \href{https://github.com/Mango0000/Pi-GS}{https://github.com/Mango0000/Pi-GS}.
\section{Related Work}

This section reviews prior work on 3D reconstruction, covering classical geometry-based pipelines, neural radiance fields,
and Gaussian splatting approaches, with a focus on sparse-view and pose-free scenarios.

\subsection{Traditional 3D Reconstruction}
Classical 3D reconstruction pipelines typically rely on Structure-from-Motion (SfM) to achieve camera pose estimation
and to generate a point cloud from a given set of images taken from various viewpoints.
Afterward, Multi-View Stereo (MVS) and surface reconstruction techniques such as Poisson reconstruction are used~\cite{NeRF-Analysis, SFM-Survey, kazhdan2006poisson}.
These methods perform well in textured and opaque scenes but struggle with transparent materials and sparse or low-overlap views.
Moreover, they are highly sensitive to SfM failures, which can lead to unstable surface reconstruction.

\subsection{Neural Radiance Fields}
Neural Radiance Fields (NeRF)~\cite{NeRFOriginal} represent scenes by continuous volumetric functions.
This makes them capable of producing photorealistic novel views and handling view-dependent effects more accurately.
However, a downside is that NeRFs are quite demanding in terms of computation.
As a result, we see more efficient variants like \textit{Instant-NGP}~\cite{mueller2022instantngp},
\textit{PlenOctree}~\cite{PlenOctree} and \textit{EfficientNeRF}~\cite{EfficientNerf} that drastically shorten
the training and rendering time by incorporating optimized data structures and improving the architecture.

\subsection{3D Gaussian Splatting}
\textit{3D Gaussian Splatting} (3DGS)~\cite{3dgs2023} has emerged as a new method
that improves on training and rendering speed by replacing the implicit radiance field
of NeRF-based methods with an explicit representation. Its core idea is to use 3D Gaussians
both for optimization and for rendering via rasterization, therefore achieving real-time rendering
without losing either fine details or transparency. Advanced 3DGS methods, such as
\textit{PGSR: Planar-based Gaussian Splatting for Efficient and High-Fidelity Surface Reconstruction}
(PGSR)~\cite{pgsr2024}, improve Gaussian surface alignment with the help of planar Gaussians and multi-view
consistency losses.
However, these methods generally rely on SfM for initialization and are optimized for dense and overlapping views.

\subsection{Sparse-View Gaussian Splatting}
Reconstruction from sparse views remains a major challenge for 3DGS.
Several augmentations exist that address sparse-view reconstruction by introducing additional constraints and regularization terms.
Depth-based supervision is explored in \textit{Depth-Regularized 3D Gaussian Splatting}~\cite{depthreg3dgs},
\textit{Few-shot NVS with Depth-Aware 3D Gaussian Splatting}~\cite{few-shotNVS3DGS}, and
\textit{DNGaussian}~\cite{DNGaussian}. This type of supervision results in fast convergence and reduces depth ambiguities.
Meanwhile, \textit{DropGaussian}~\cite{DropGaussian} and \textit{DropoutGS}~\cite{DropoutGS}
deactivate Gaussians at random in order to counteract overfitting.
There are also more advanced methods like FSGS: Real-Time Few-Shot View Synthesis using Gaussian Splatting~\cite{fsgs} which
introduces a pooling strategy and fine-tunes the splitting strategy to improve sparse view reconstruction across different datasets.
While these methods achieve very robust results in sparse-view scenarios, they typically rely on accurate camera poses from SfM.

\subsection{SfM-Free Methods}
Methods such as \textit{COLMAP-Free 3D Gaussian Splatting}~\cite{COLMAPFREE3DGS}
and \textit{InstantSplat}~\cite{InstantSplat}, eliminate the need for SfM by jointly optimizing the 3D Gaussians
as well as the camera poses and using depth estimations for point cloud initialization. These methods are able to
handle sparse-view situations more robustly and recover from inaccurate camera poses.

\subsection{Diffusion-Based Priors}
More recent works incorporate diffusion priors not only to stabilize the reconstruction,
but also to generate additional views from the limited number of input views.
\textit{GenFusion}~\cite{GenFusion}, \textit{SparseGS}~\cite{SparseGS},
\textit{Gaussian Scenes}~\cite{GaussianScenes}, and \textit{Intern-GS}~\cite{InternGS}
are some of the methods where these advantages can also be observed.
While these methods achieve impressive results, they often struggle with high-frequency textures
and view inconsistencies due to depth ambiguities and inaccurate Gaussian alignment.
\\\\
Our method differs fundamentally from diffusion-based and optimization-heavy approaches.
Instead of synthesizing novel views using generative priors, we improve reconstruction
quality through dense geometric initialization and strong generalizability across datasets.
We leverage depth and normal supervision from estimated depth maps and explicitly model depth
uncertainty through confidence-aware constraints, allowing deviations from noisy estimates. Camera poses
and point representations are predicted by a feed-forward network, reducing reliance on iterative optimization
and increasing robustness in sparse-view settings. Consequently, our approach focuses on geometric consistency and
generalization without relying on view hallucination or diffusion-based priors.
\section{Method}
We begin by outlining preliminaries on Gaussian Splatting and planar depth rendering. \Cref{sec:pgsr_modifications} details modifications to PGSR for sparse settings, followed by our dense initialization strategy in \Cref{sec:dense_initialization}. We then present our uncertainty-aware Pearson loss in \Cref{sec:depth_supervision} and artifact-free normal supervision in \Cref{sec:normal_supervision}. Finally, \Cref{sec:depth_warping} describes our depth warping approach for improving view consistency.

\subsection{Preliminaries}

\paragraph{Gaussian Splatting.}
\textit{3D Gaussian Splatting} (3DGS) introduced by Kerbl et al.~\cite{3dgs2023} achieves great novel view synthesis results with high efficiency by leveraging a Gaussian scene representation. Another improvement of this scene representation over NeRF is the real-time rendering speed, as well as much faster training times. Our approach also builds upon 3DGS.
The scene representation is defined by a set of 3D Gaussians.
Each Gaussian can be defined by a 3D covariance matrix $\Sigma \in \mathbb{R}^{3 \times 3}$ and the 3D center point $\mu \in \mathbb{R}^3$ in world space,
\begin{equation}
    G(x) = e^{-\frac{1}{2}(x - \mu_i)^T\Sigma^{-1}(x - \mu_i)}.
\end{equation}

To project this 3D Gaussian onto the 2D image plane for rendering, the covariance matrix $\Sigma'$ in clip space is defined as the following:
\begin{equation}
    \Sigma' = J W \Sigma W^T J^T,
\end{equation}
where $J$ is the Jacobian of the affine approximation for this projection transformation and $W$ is the view transformation matrix.

For the covariance matrix to be physically meaningful, it needs to be positive semi-definite.
To ensure this throughout the training process, $\Sigma$ is defined as the following:
\begin{equation}
    \Sigma = R S S^T R^T,
\end{equation}
where $S \in \mathbb{R}^{3\times 3}$ is the scaling matrix, and $R \in \mathbb{R}^{3\times 3}$ is the rotation matrix.
This allows separate optimization of rotation and scaling and ensures that $\Sigma$ is positive semi-definite. For increased memory efficiency, the rotation matrix is stored as a quaternion, and scaling as 3D vector.

Furthermore, for rendering the color $C$, we blend the colors of each Gaussian along the ray, as follows:
\begin{equation}
    C = \sum^N_{i=1} T_i \alpha_i c_i,
\end{equation}
where $N$ is the number of Gaussians along a ray, $c_i$ is the color of the i-th Gaussian represented by spherical harmonics (SH) to account for view dependent effects, $\alpha_i$ is the weighted opacity of the i-th Gaussian and $T_i$ is the transmittance of the i-th Gaussian~\cite{3dgs2023}.

Transmittance $T_i$ is defined as:
\begin{equation}
\label{eq:transmittance}
    T_i = \prod^{i-1}_{j=1}(1 - \alpha_j).
\end{equation}
By calculating the color for each ray from the camera, we can render an image.
The training of this Gaussian representation is done by back propagation with the following loss function:
\begin{equation}
    \mathcal{L} = (1 - \lambda) \mathcal{L}_1 + \lambda \mathcal{L}_{D-SSIM},
\end{equation}
where $\mathcal{L}_1$ is a simple $l_1$ loss between the rendered and ground-truth image and $\mathcal{L}_{D-SSIM}$ is an image similarity measure between rendered and ground-truth image~\cite{3dgs2023, dssim}.
3DGS relies on camera poses and points obtained from structure from motion (SfM). However, in sparse-view settings, the resulting point cloud can be highly sparse, and the overlap between the images may be insufficient to extract reliable structures or accurate camera poses. This leads to a challenging starting point for 3DGS optimization.

\paragraph{Depth and Normal Rendering.}
We use \textit{Planar-based Gaussian Splatting for Efficient
and High-Fidelity Surface Reconstruction}~\cite{pgsr2024} (PGSR) for normal and depth rendering. PGSR builds upon 3DGS, enabling the rendering and backpropagation of both the depth and normals.
A naive approach of computing the depth $D$ of a pixel would be to use depth accumulation defined as:
\begin{equation}
    D = \sum_{i=1}^N T_i \alpha_i z_i,
\end{equation}
where $T_i$ is the same as in ~\cref{eq:transmittance}, $\alpha_i$ is the weighted opacity of the i-th Gaussian and $z_i$ is its distance from the camera~\cite{gaussianproDepth}.
PGSR on the other hand compresses the 3D Gaussians to get flat 2D planes, from which unbiased depth and normal maps can be rendered~\cite{pgsr2024}.

To get the 2D planes, PGSR flattens the 3D Gaussians by minimizing the minimum scale and therefore defining the scale loss $\mathcal{L}_s$ as following:
\begin{equation}
    \mathcal{L}_s = ||\min(s_1, s_2, s_3)||_1,
\end{equation}
where $s_i$ is the i-th scale component of each Gaussian.

The direction of the minimum scale factor corresponds to the normal $n_i$.
Therefore, the normals per ray, $\mathcal{N}$, can be rendered as following:
\begin{equation}
    \mathcal{N} = \sum_{i=1}^N R_c^T n_i \alpha_i T_i,
\end{equation}
where $R_c$ is the rotation from the camera to the global world.

The distance $d_i$ from the Gaussian plane to the camera center is defined as:
\begin{equation}
    d_i = (R_c^T (\mu_i - T_c)) R_c^T n_i^T,
\end{equation}
where $T_c$ is the camera center in the world and $\mu_i$ is the center of the i-th Gaussian.

The distance $D$ along a ray can now be defined as:
\begin{equation}
    D = \sum_{i=1}^N d_i \alpha_i T_i.
\end{equation}
PGSR extends 3DGS by introducing an Image Edge-Aware Single-View Loss $\mathcal{L}_{svgeo}$, which optimizes the Gaussian Scene with the Local Plane Assumption. This assumption states that two neighbouring pixels can be considered as an approximate local plane, but only if these pixels do not belong to an edge. The loss helps to improve the local depth and normal consistency. They also propose a Multi-View Geometric Consistency Loss, $\mathcal{L}_{mvgeom}$, which enhances geometric smoothness by projecting the depth and normals from one frame to another. Finally, they employ a Multi-View Photometric Consistency Loss, $\mathcal{L}_{mvrgb}$, which projects the grayscale image from one camera to another camera through depth warping~\cite{pgsr2024}.

\subsection{PGSR Sparse-View}
\label{sec:pgsr_modifications}
\begin{figure}
\centering
\begin{subfigure}{.24\textwidth}
    \centering
    \captionsetup{justification=centering, width=0.9\textwidth}
    \includegraphics[width=0.98\linewidth]{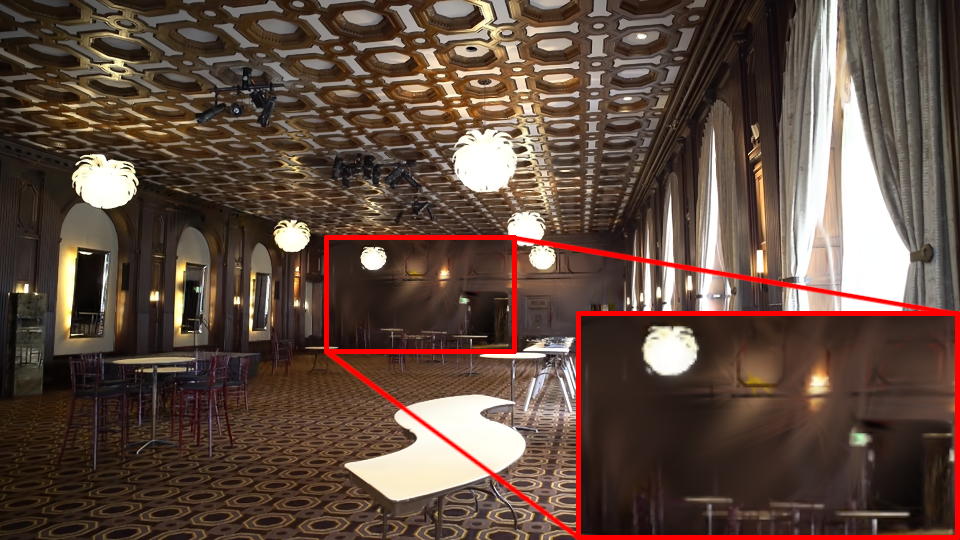}
    \caption{Ballroom scene with opacity reset.}
    \label{fig:opacity_reset}
\end{subfigure}%
\begin{subfigure}{.24\textwidth}
    \centering
    \captionsetup{justification=centering, width=0.9\textwidth}
    \includegraphics[width=0.98\linewidth]{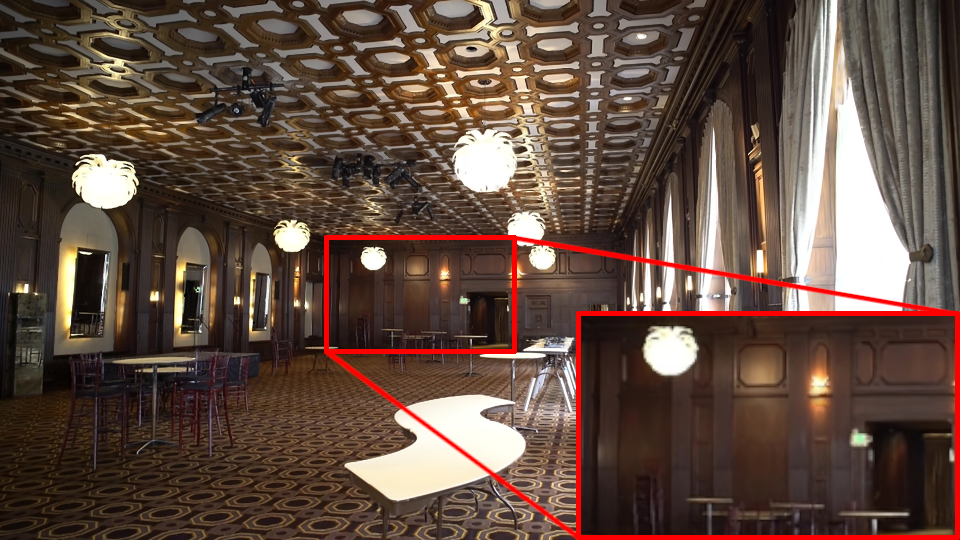}
    \caption{Ballroom scene without opacity reset.}
    \label{fig:no_opacity_reset}
\end{subfigure}
\caption{Comparison of the Ballroom scene from Tanks and Temples with and without opacity reset~\cite{tankstemples}. Background details are lost when opacity reset is executed, and image quality further degrades over the training process.}
\label{fig:opacity_reset_example}
\end{figure}

Default PGSR does not work well for the sparse-view setting out-of-the-box because of the multi-view observer trim, which assures that each point is observed by multiple cameras and this is not guaranteed in sparse-view settings. Therefore, we deactivate this trimming for our method.
Another parameter that requires adjustment is the opacity reset interval. When opacity reset happens, fine details in the background will be lost and artifacts appear, as can be seen in \cref{fig:opacity_reset_example}. The details in \cref{fig:opacity_reset} at the back wall are completely lost and artifacts in the window frame become visible. By continuing the training process even further, the artifacts' strength increases, and they become even more prominent. When deactivating opacity reset, the background details are retained and the artifacts vanish without sacrificing the overall quality. This can also be seen in \cref{fig:no_opacity_reset}. The improvement is also reflected in the PSNR (Peak Signal-to-Noise Ratio), which increases from \textit{22.76} to \textit{23.73}.
With these few settings, it is already possible to run the PGSR framework with acceptable results. For improved performance, we deactivate the splitting strategy as it is not needed for our dense point cloud initialization. The point cloud is already very detailed and this setting does not improve the final results (\textit{cf.} \cref{tab:ablation}).

\subsection{Dense Initialization}
\label{sec:dense_initialization}
\begin{figure}
\centering
\begin{subfigure}[t]{.25\textwidth}
  \centering
  \captionsetup{justification=centering}
  \includegraphics[width=0.98\linewidth]{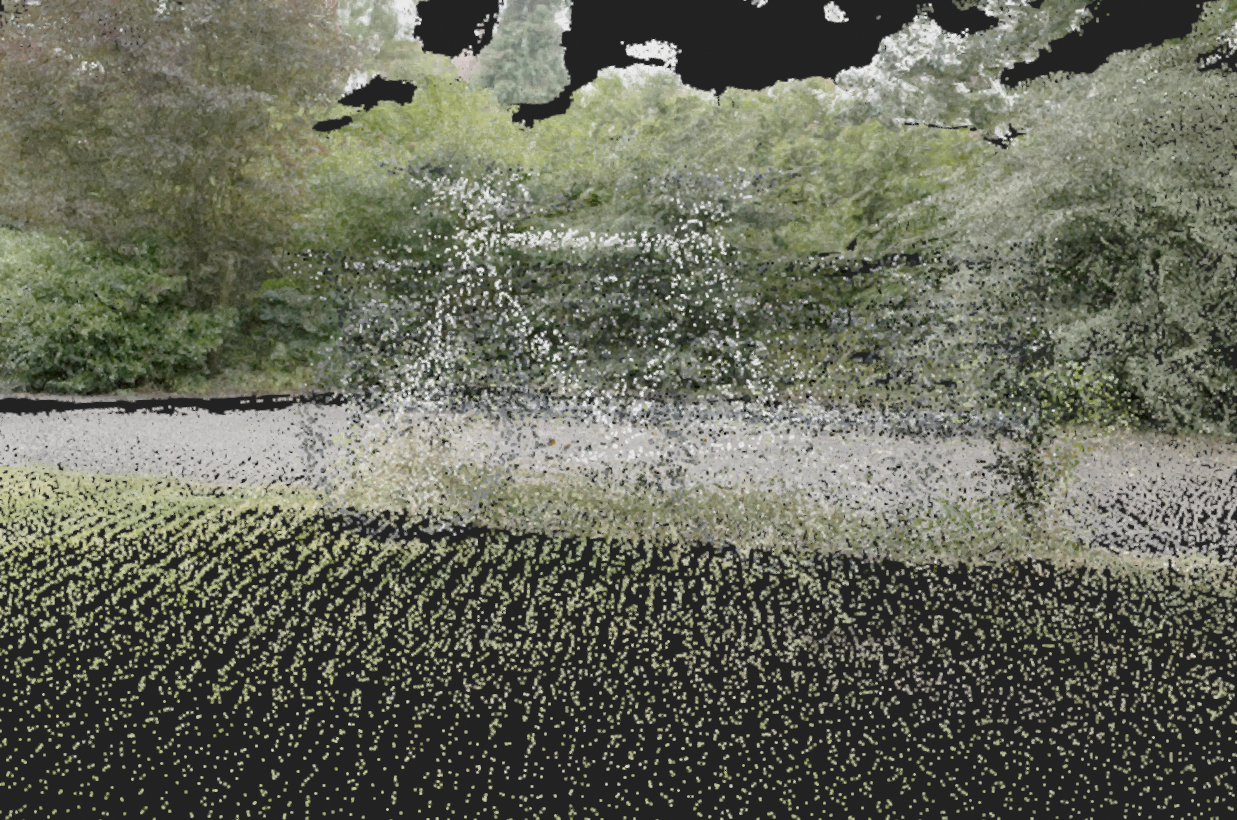}
  \caption{Point cloud inferred with $\pi^3$.}
  \label{fig:pointcloud_pi3}
\end{subfigure}%
\begin{subfigure}[t]{.25\textwidth}
  \centering
  \captionsetup{justification=centering}
  \includegraphics[width=0.98\linewidth]{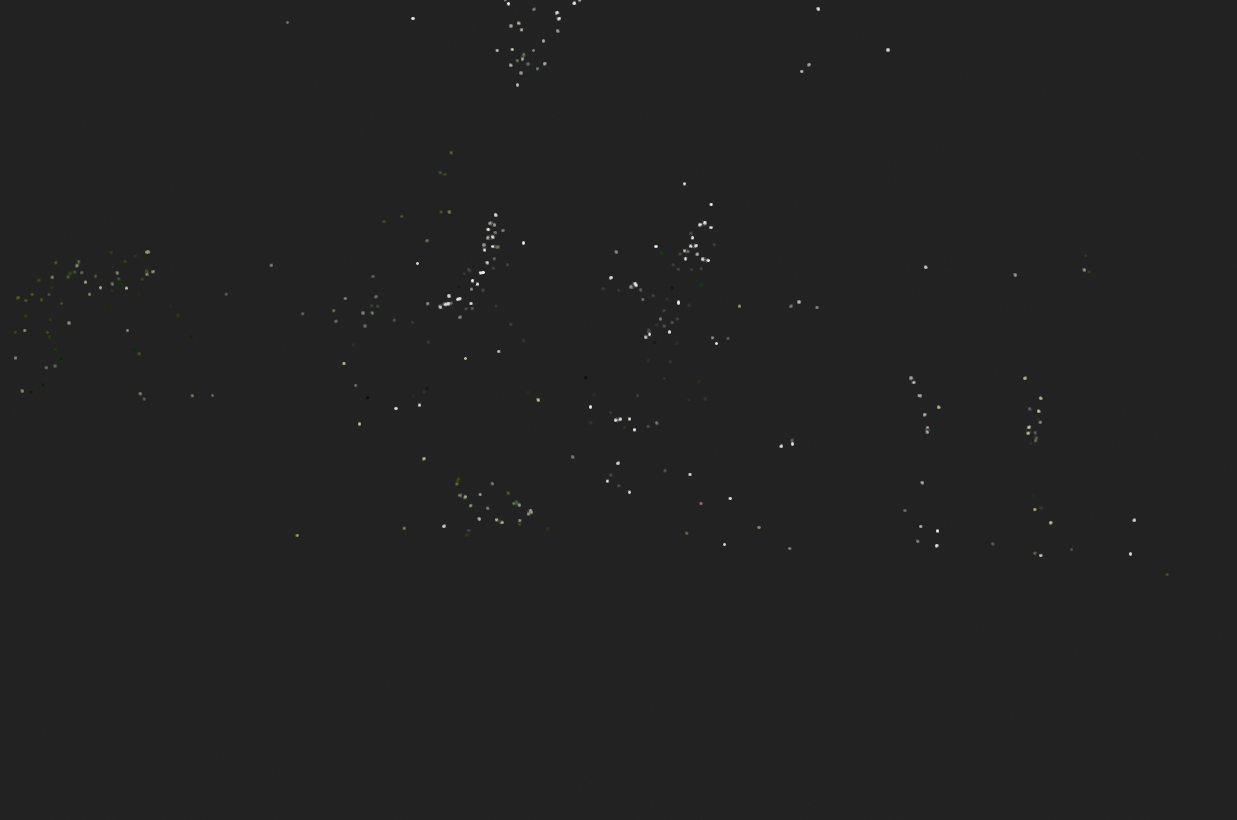}
  \caption{Point cloud created with COLMAP.}
  \label{fig:pointcloud_colmap}
\end{subfigure}
\caption{Comparison between $\pi^3$ point cloud and COLMAP point cloud, of the bike scene from MipNeRF360 Dataset with 24 training images~\cite{mipnerf360,wang2025pi3,SFM-Original}.}
\label{fig:pointcloud_comparison}
\end{figure}

Sparse-view settings pose a fundamental challenge for standard SfM frameworks like COLMAP~\cite{SFM-Original, MVSPixel}, where limited image overlap can lead registration to fail. Furthermore, the resulting sparse point clouds serve as a poor initialization for 3DGS, complicating the optimization of Gaussian primitives and compromising geometric fidelity. To mitigate this, we leverage a pre-trained feed-forward network to predict both depth and camera parameters. This strategy provides the dense geometric initialization and accurate poses required for high-quality sparse-view reconstruction.
\Cref{fig:pointcloud_pi3} illustrates the point cloud generated by the feed-forward model $\pi^3$~\cite{wang2025pi3}, while \cref{fig:pointcloud_colmap} depicts the result from COLMAP~\cite{SFM-Original}. Both methods use the same 24 input views from the "bicycle" scene of the MipNeRF360 dataset~\cite{mipnerf360}, rendered here from an identical viewpoint. The difference in density is significant: The COLMAP reconstruction contains only 1,028 points, whereas $\pi^3$ yields 1,013,106 points. Note that the $\pi^3$ output was filtered using the default confidence threshold of 20\%.

\subsection{Depth Supervision}
\label{sec:depth_supervision}
From $\pi^3$, we obtain the per view point clouds which can be used as a depth map.
For depth regularization, we evaluated different losses.

Standard L1 and L2 losses often cause the model to overfit to the limited fidelity of the inferred depth maps. We also evaluated the Global-Local Depth Normalization from \textit{DNGaussian}~\cite{DNGaussian} but found it unnecessary given the inherent scale consistency of our predictions. Instead, we utilize a Pearson correlation loss, which has demonstrated superior performance. This approach enforces structural consistency while enabling the recovery of high-frequency details that are missing from the initial depth estimation.

\begin{figure}
\centering
\begin{subfigure}{.25\textwidth}
    \centering
    \captionsetup{justification=centering}
    \includegraphics[width=.98\linewidth]{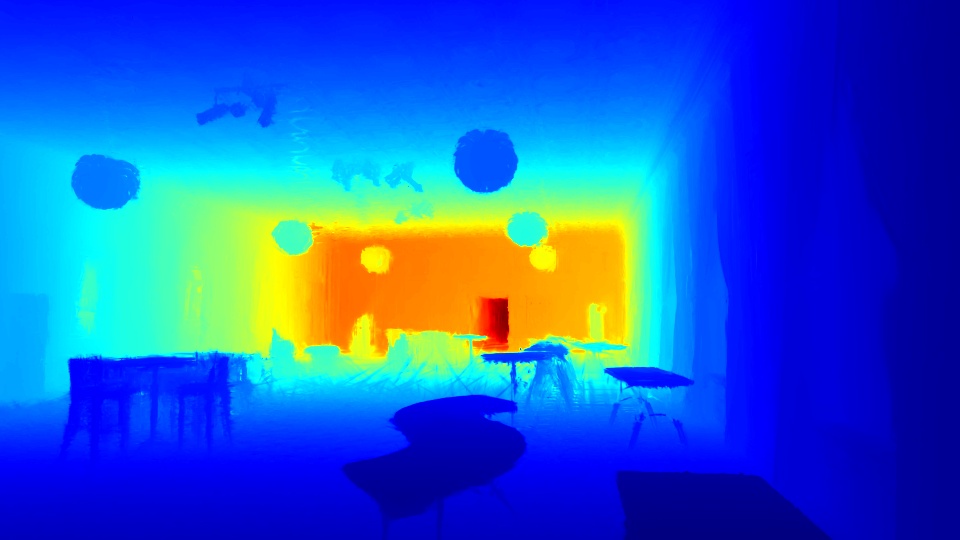}
    \caption{Confidence-aware Peason loss.}
    \label{fig:conf_aware_pearson_depth}
\end{subfigure}%
\begin{subfigure}{.25\textwidth}
    \centering
    \captionsetup{justification=centering}
    \includegraphics[width=.98\linewidth]{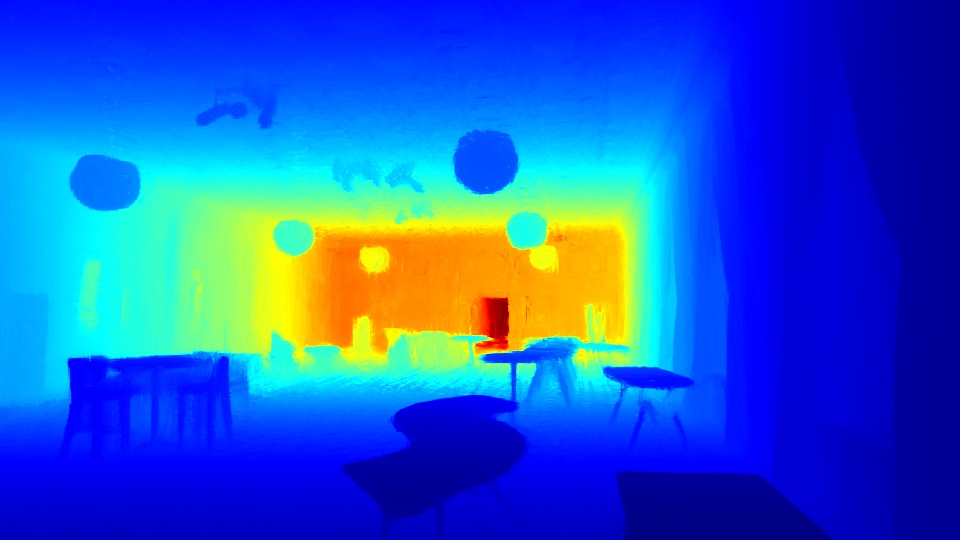}
    \caption{Pearson loss.}
    \label{fig:standard_pearson_depth}
\end{subfigure}
\caption{Depth rendering of the Ballroom scene from the Tanks and Temples dataset, comparing the confidence-aware Pearson loss with the standard pearson loss~\cite{tankstemples}. The confidence-aware loss leverages uncertainty estimates to enhance detail, particularly in the background, and also improves performance with low-resolution depth estimates.}
\label{fig:PCC_Ballroom}
\end{figure}

In addition to the default Pearson correlation loss, we also integrated the confidence given by $\pi^3$. As a result, the final depth can be modeled even more accurately by assigning low weights to uncertain regions.
Our newly created confidence-aware depth loss, $\mathcal{L}_{pearson}$, is defined as:
\begin{equation}
    \mu_p = \frac{\sum^N_{i=1} C_i D^p_{i}}{\sum^N_{i=1} C_i}, \quad \mu_t = \frac{\sum^N_{i=1} C_i D^t_{i}}{\sum^N_{i=1} C_i},
\end{equation}
\begin{equation}
    \bar{D}_p = D_p - \mu_p, \quad \bar{D}_t = D_t - \mu_t,
\end{equation}
\begin{equation}
    \small
    P_{conf} = \frac{\sum^N_{i=1} C_i \bar{D}^p_{i} \bar{D}^t_{i}}{\sqrt{\left(\sum^N_{i=1} C_i \left(\bar{D}^p_{i}\right)^2\right)\left(\sum^N_{i=1} C_i \left(\bar{D}^t_{i}\right)^2\right)} },
\end{equation}
\begin{equation}
    \mathcal{L}_{pearson} = 1- P_{conf},
\end{equation}
$N$ is the number of pixels, $D^p_{i}$ is the predicted depth of the i-th pixel, $C_i$ the confidence of the i-th pixel and $D^t_{i}$ is the ground truth of the i-th pixel, which is the depth estimated by $\pi^3$, and $P_{conf}$ is the confidence-aware Pearson correlation.
The resulting rendered depth after 7,000 iterations with the help of confidence-aware Pearson correlation can be seen in \cref{fig:PCC_Ballroom}.



\subsection{Normal Supervision}
\label{sec:normal_supervision}
\begin{figure}
\centering
\begin{subfigure}{.25\textwidth}
    \centering
    \captionsetup{justification=centering}
    \includegraphics[width=.98\linewidth]{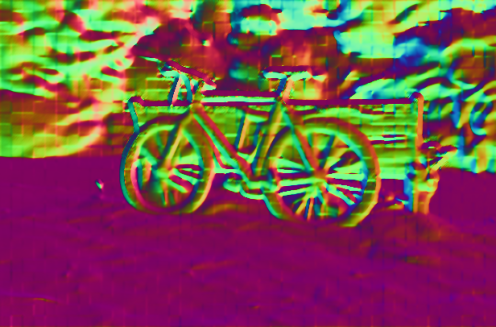}
    \caption{Default normal map with artifacts.}
    \label{fig:normal_artifacts}
\end{subfigure}%
\begin{subfigure}{.25\textwidth}
    \centering
    \captionsetup{justification=centering}
    \includegraphics[width=.98\linewidth]{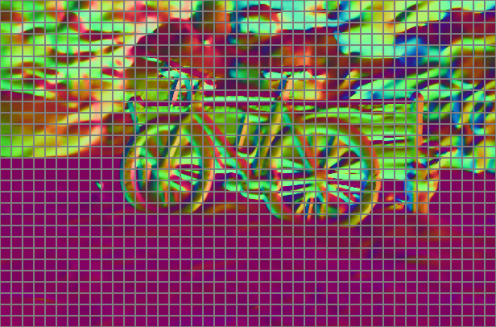}
    \caption{Masked normal map.}
    \label{fig:normal_masked}
\end{subfigure}
\caption{The Normal map generated from the depth map using partial derivatives, which introduces grid artifacts, and the masked normal map, which removes grid artifacts introduced by $\pi^3$ architecture~\cite{wang2025pi3}.}
\label{fig:normals}
\end{figure}

Surface Normals can be computed with the help of depth maps by calculating the pixel-wise partial derivatives 
$\frac{\partial z}{\partial x}$ and $\frac{\partial z}{\partial y}$, where $x$ and $y$ are the pixel coordinates and $z$ is the depth value, either rendered or estimated by $\pi^3$.
Because $\pi^3$ processes each image in patches of $14 \times 14$ pixels, the gradient is not continuous between adjacent patches, leading to grid-like artifacts, as can be seen in \cref{fig:normal_artifacts}.
To alleviate this problem, we add a mask to ignore these discontinuous regions during loss computation.
The mask is computed by creating a grid with $14 \times 14$ pixel cells, masking the 1-pixel-wide inner border of each cell.
Therefore, the Gaussians are not regularized in these border regions, and the grid artifacts do not appear in the scene representation. The masked normal map can be seen in \cref{fig:normal_masked}.
As supervision, we simply use the L1 loss between the rendered and ground-truth normal map defined as:
\begin{equation}
    \mathcal{L}_{normal}=\frac{1}{N}\sum^N_{i=1}||N^t_{i}-N^p_{i}||_1,
\end{equation}
where N is the number of pixels, $N^t_{i}$ is the ground-truth normal at pixel i and $N^p_{i}$ is the predicted normal at pixel i.

\subsection{Depth Warping}
\label{sec:depth_warping}

\begin{figure}
\centering
\begin{subfigure}{.25\textwidth}
    \centering
    \captionsetup{justification=centering}
    \includegraphics[width=.98\linewidth]{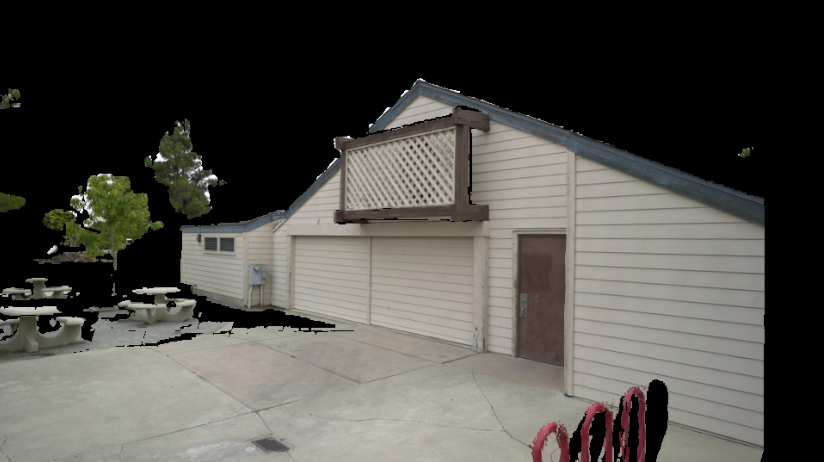}
    \label{fig:reprojection_1}
\end{subfigure}%
\begin{subfigure}{.25\textwidth}
    \centering
    \captionsetup{justification=centering}
    \includegraphics[width=.98\linewidth]{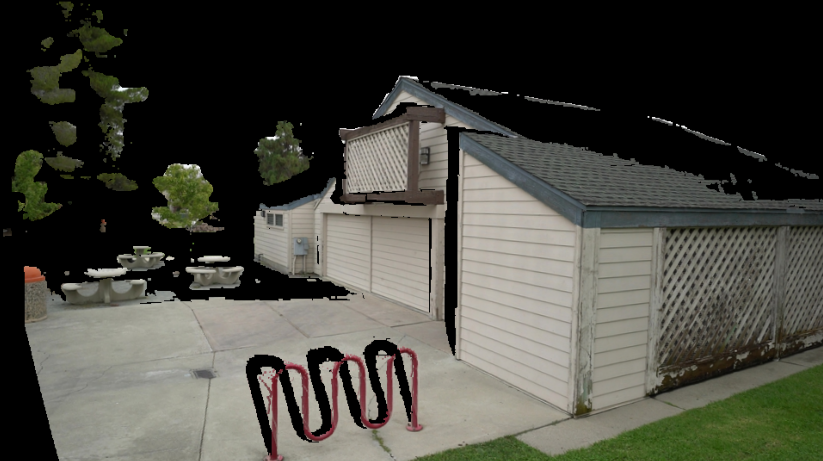}
    \label{fig:reprojection_2}
\end{subfigure}
\caption{The two Figures show two reprojection examples with the applied mask for the Barn scene of the Tanks and Temples dataset~\cite{tankstemples}.}
\label{fig:reprojection}
\end{figure}

To improve generalization of our model further, we include pseudo-views which are generated with the help of depth warping. This is achieved by projecting the image pixels from one camera into 3D space, and then reprojecting the 3D points into the 2D image plane of a target camera.
For accurate results, we only project pixels with high confidence and mask out the rest, including unseen regions. To generate high-quality pseudo-cameras, we use circle interpolation with the camera parameters as input. A circle can be defined by three points, so we use the two nearest cameras to the target camera for pseudo-view generation. The positions of the three cameras define our circle. Now then interpolate by a certain amount between each pair of neighbouring views, which results in two additional views per camera. We can generate an arbitrary number of pseudo-views by adjusting the interpolation step size. However, in our experiments, two pseudo-views between each pair yielded the best results.
The nearest cameras are already computed by PGSR, therefore we can reuse them. A few examples of these generated pseudo-views can be seen in \cref{fig:reprojection}.
These pseudo-views are then used throughout training for additional supervision with the help of SSIM and L1 loss, but with a weight set to \textit{0.1}.
\section{Evaluation}

For testing strategy, we adhere to previous state-of-the-art models to ensure comparability.
The datasets used for the evaluation are Tanks and Temples~\cite{tankstemples}, MipNeRF360~\cite{mipnerf360}, LLFF~\cite{mildenhall2019llff} and DTU~\cite{dtu}.

\paragraph{Implementation Details.}
The Tanks and Temples dataset covers real-world indoor and outdoor scenes, but we only use a subset of 8 scenes, as done by other sparse-view models like \textit{Intern-GS} and \textit{InstantSplat}. We focus on the 3-view setting and therefore use the same train/test split. This means the testing set includes 12 images uniformly sampled without the first and last frame and the remaining set is the training set where we again uniformly sampled the 3 views~\cite{InstantSplat}. For Tanks and Temples, no downsampling is applied.

The MipNeRF360 dataset contains real-world $360^\circ$ indoor and outdoor scenes. For this dataset, two different approaches are used. One for the 3-view setting as defined by \textit{Gaussian Scenes}~\cite{GaussianScenes} and one for the 12-view setting as defined by \textit{SparseGS}~\cite{SparseGS}. For both settings, the 4x downsampled images are used, to adhere to the evaluation strategies of state-of-the-art models.  For the 3-view setting, we use every 8th image as testing set and uniformly sample the 3 training views. For the 12-view setting, we use the split dataset provided by \textit{SparseGS}~\cite{SparseGS}. The 12-view setting uses only 6 of the 9 scenes contained in the MipNeRF360 dataset, whereas the 3-view setting uses all 9 scenes.

The LLFF dataset contains real-world forward-facing images. For this dataset, we used the same evaluation strategy as defined by \textit{DNGaussian}. A downsampling rate of 8 is used, and we adhere to the train/test split of the 3-view setting of \textit{DNGaussian}~\cite{DNGaussian}.

Lastly, we also evaluated on the DTU dataset, which contains highly calibrated lab captures of object centric scenes. This dataset also provides bit masks to separate the background and real camera poses. We used our own inferred camera poses. We again used the testing strategy defined by \textit{DNGaussian}. This time we used 4x downsampled images and the same train/test split of the 3-view setting of \textit{DNGaussian}~\cite{DNGaussian}. Similar to DNGaussian and other comparable methods, we applied the provided separation masks for the evaluation.

We use the exact same settings for all evaluations. $\pi^3$~\cite{wang2025pi3} automatically downsamples the images to a certain pixel size, therefore we counteract the downsampling by rescaling the cameras to the full size. To make a fair comparison, we only project the training views to 3D space. The testing views are only used to get initial camera positions.
We train for 7000 iterations, with depth loss, normal loss as well as pseudo views. The pseudo views are generated with a confidence threshold of 20\%. This means that we mask out the projected pixel with confidence under 20\%. Splitting of Gaussians is deactivated.
We evaluate our model in terms of PSNR, SSIM and LPIPS.

\subsection{Quantitative Evaluation}
\begin{figure}
\centering
\begin{subfigure}{.25\textwidth}
    \centering
    \captionsetup{justification=centering}
    \includegraphics[width=.96\linewidth]{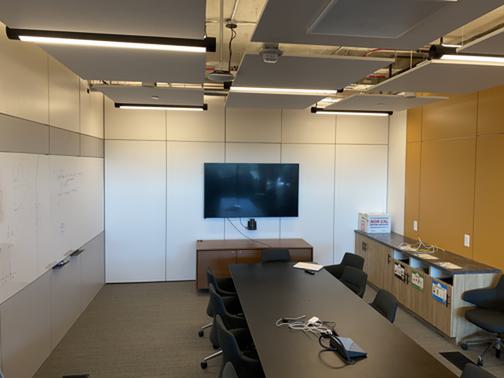}
    \caption{Ground Truth}
    \label{fig:llff_gt}
\end{subfigure}%
\begin{subfigure}{.25\textwidth}
    \centering
    \captionsetup{justification=centering}
    \includegraphics[width=.96\linewidth]{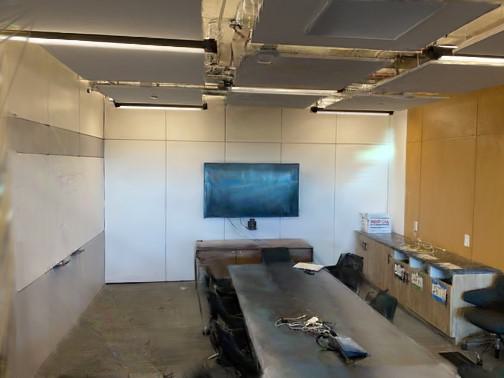}
    \caption{Intern-GS}
    \label{fig:llff_interngs}
\end{subfigure}
\vspace{0.3cm}
\begin{subfigure}{.25\textwidth}
    \centering
    \captionsetup{justification=centering}
    \includegraphics[width=.96\linewidth]{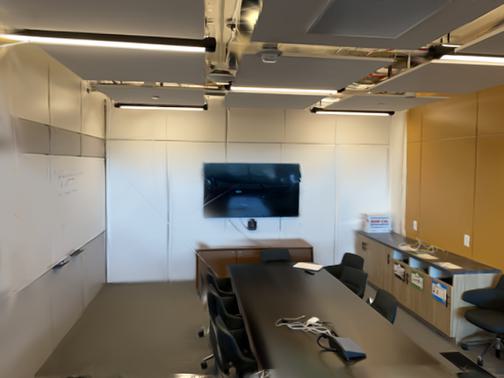}
    \caption{Ours}
    \label{fig:llff_ours}
\end{subfigure}
\caption{The Figures show a comparison between Intern-GS~\cite{InternGS}, Ours and the Ground Truth. Our model has very accurate reflections and fewer artifacts, nevertheless our model can not correctly reconstruct the unseen region at the ceiling.}
\label{fig:llff_example}
\end{figure}

\Cref{tab:llff_dtu_comparison,tab:evaluation_tat} show the comparison between \textit{Intern-GS}~\cite{InternGS}, \textit{InstantSplat}~\cite{InstantSplat}, \textit{SparseGS}~\cite{SparseGS}, \textit{DNGaussian}~\cite{DNGaussian}, \textit{FSGS}~\cite{fsgs}, 3DGS~\cite{3dgs2023} and Our method. On DTU and Tanks and Temples, our model can reconstruct the scene accurately, with good Gaussian surface alignment and without smoothing out high-frequency textures. On LLFF our model achieves slightly lower scores, because of missing information in unseen regions, as our model optimizes only on seen regions and known information. An example of this unseen region is illustrated in \cref{fig:llff_example}.

\cref{tab:evaluation_mipnerf360_3view} shows the comparison between Gaussian Scenes, MASt3R Initialization, FSGS and Our method in the 3-view setting on MipNeRF360~\cite{GaussianScenes, fsgs}. Our model achieves the lowest LPIPS score and second highest PSNR and SSIM. Compared to FSGS our model does not rely on accurate camera poses from traditional SfM.

\begin{figure}
\centering
\begin{subfigure}{.25\textwidth}
    \centering
    \captionsetup{justification=centering}
    \includegraphics[width=.96\linewidth]{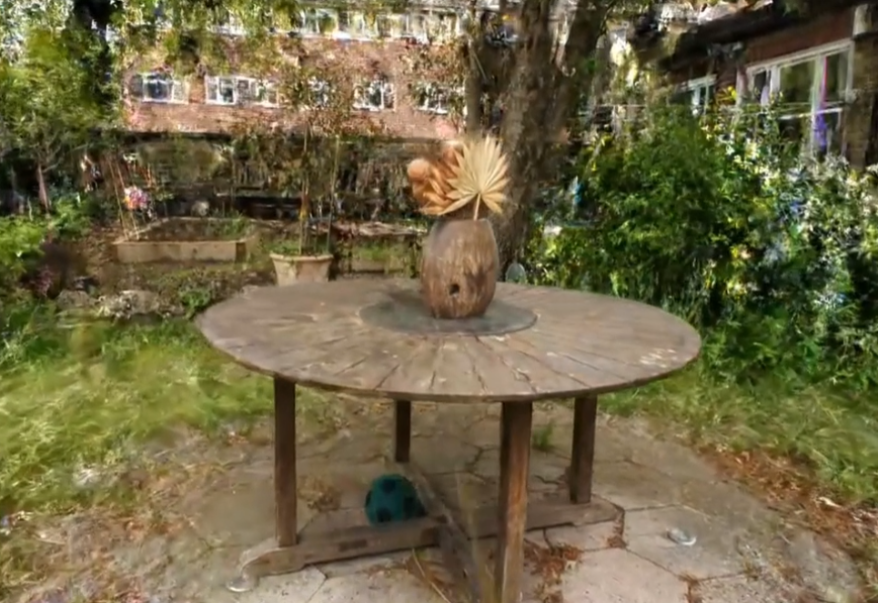}
    \caption{SparseGS}
    \label{fig:garden_sparsegs}
\end{subfigure}%
\begin{subfigure}{.25\textwidth}
    \centering
    \captionsetup{justification=centering}
    \includegraphics[width=.96\linewidth]{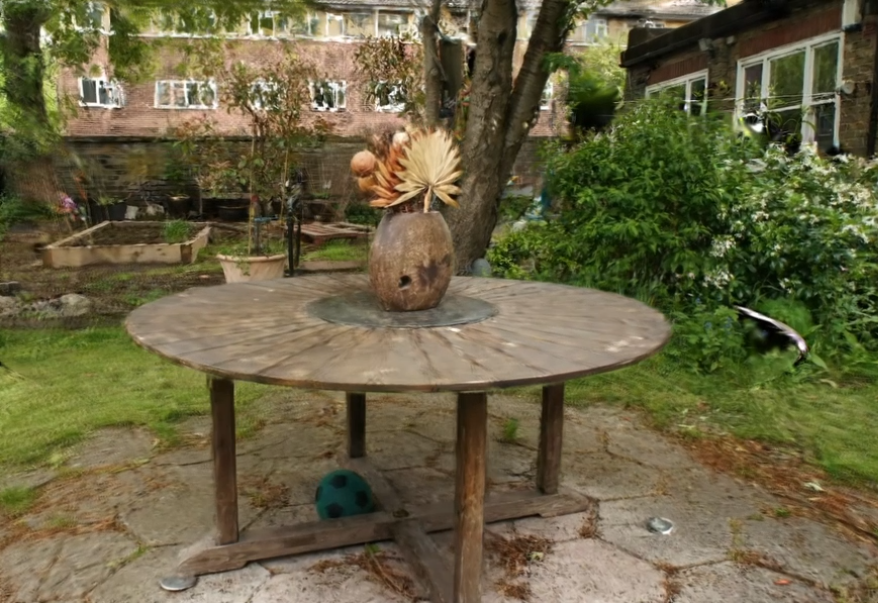}
    \caption{Ours}
    \label{fig:garden_ours}
\end{subfigure}
\caption{The Figures show a comparison between SparseGS~\cite{SparseGS} and our method. Our model reconstructs the background and ground more accurately, and additionally decreases artifacts.}
\label{fig:garden_example}
\end{figure}

\cref{tab:evaluation_mipnerf360_12view} shows the comparison between 3DGS, \textit{DNGaussian}, \textit{SparseGS} and Our method in 12-view setting on MipNeRF360~\cite{mipnerf360}. Our model achieves the highest results with very coherent and view-consistent final scene, as our model improves the Gaussian surface alignment significantly. A comparison can be seen in \cref{fig:garden_example}.\\
To validate the accuracy of our camera pose estimates, we evaluate the Absolute Trajectory Error (ATE) on the Tank and Temples dataset. Our pose estimator, $\pi^3$, achieves a mean ATE of 0.0293 and a root mean squared error (RMSE) of 0.0325, demonstrating that it produces accurate camera poses suitable for fair comparison of photometric metrics in 3D Gaussian splatting.

\subsection{Ablation}
We evaluate the impact of each individual optimization on our final result.
The evaluation is conducted using the Barn scene from the Tanks and Temples dataset. It is evident that all of our optimizations improve the result even further. Dense point cloud initialization with the help of $\pi^3$ significantly improves the result by also reducing the time required for SfM.
Our custom depth loss improves the score by allowing low confidence depth regions to optimize more freely.
Normal regularization encourages the Gaussians' normals to match the ground-truth geometry. Depth warping improves the results by adding more views, which helps the model generalize better and avoid overfitting to the training views.
Our full model achieves a PSNR of \textit{22.15} on the Barn scene. We also evaluated the effect of enabling splitting of Gaussians in our model. This setting results in a slight decrease in performance and was therefore deactivated. These results can be seen in \cref{tab:ablation}.
\begin{table}[ht!]
    \centering
    \footnotesize
    \setlength{\tabcolsep}{2pt}
    \begin{tabular}{lc}
    \toprule
     Method & PSNR\\
     \midrule
 Original 3DGS & 17.53\\
 PGSR & 18.05\\
 $\pi^3$ (dense) initialization & 19.66\\
 + Depth Regularization & 20.72\\
 + Normal Regularization & \cellcolor{tab_color!10}21.56\\
 + Depth Warping (Full Model) & \cellcolor{tab_color!50}{22.15}\\
 \midrule
 + Splitting Densification &\cellcolor{tab_color!30}21.97\\
    \bottomrule
    \end{tabular}
    \caption{Ablation study of the regularization techniques introduced in our model on the Barn scene of Tanks and Temples. Additionally, we evaluated the impact of splitting Gaussians during densification.}
    \label{tab:ablation}
\end{table}

In addition, we evaluate the impact of using PGSR compared to standard 3DGS for our sparse view setting (3-views).
\Cref{tab:evaluation_3dgs_pgsr} shows that the planar depth created by PGSR helps significantly to place the Gaussians more accurately. Additionally, the losses introduced by PGSR help to improve the rendering results further. Our model remains stable even after increased training iterations and continues to show improved novel view synthesis results.
A visual comparison between 3DGS and PGSR with different number of iterations can be seen in \cref{fig:pgsr_vs_3dgs}.
\begin{table}[ht!]
    \centering
    \footnotesize
    \setlength{\tabcolsep}{2pt}
    \resizebox{\columnwidth}{!}{\begin{tabular}{lccccccc}
    \toprule
    \multirow{3}{*}{Framework} & \multirow{3}{*}{Iteration} &\multicolumn{3}{c}{Tanks and Temples} & \multicolumn{3}{c}{MipNeRF360}\\
    \cmidrule(lr){3-5} \cmidrule(lr){6-8}
    && PSNR\textsuperscript{$\uparrow$} & SSIM\textsuperscript{$\uparrow$} & LPIPS\textsuperscript{$\downarrow$}& PSNR\textsuperscript{$\uparrow$} & SSIM\textsuperscript{$\uparrow$} & LPIPS\textsuperscript{$\downarrow$}\\
    \midrule
    PGSR~\cite{pgsr2024} & 7000  & \cellcolor{tab_color!30}19.99 & \cellcolor{tab_color!30}0.503 & \cellcolor{tab_color!30}0.355 & \cellcolor{tab_color!30}23.36 & \cellcolor{tab_color!30}0.791 & \cellcolor{tab_color!50}0.156\\
    3DGS~\cite{3dgs2023} & 7000  & \cellcolor{tab_color!10}18.00 & \cellcolor{tab_color!10}0.426 & \cellcolor{tab_color!10}0.449 & \cellcolor{tab_color!10}23.07 & \cellcolor{tab_color!10}0.773 & \cellcolor{tab_color!10}0.172\\
    \midrule
    PGSR~\cite{pgsr2024} & 15000 & \cellcolor{tab_color!50}20.19 & \cellcolor{tab_color!50}0.517 & \cellcolor{tab_color!50}0.343 & \cellcolor{tab_color!50}23.41 & \cellcolor{tab_color!50}0.795 & \cellcolor{tab_color!30}0.169\\
    3DGS~\cite{3dgs2023} & 15000 & 17.04 & 0.391 & 0.465 & 20.94 & 0.719 & 0.244\\
    \bottomrule
    \end{tabular}}
    \caption{Ablation study on the use of PGSR as base framework compared to 3DGS. The additional multi-view and single-view losses introduced by PGSR are activated after iteration 7000. This comparison shows that the PGSR depth rendering captures the underlying surface geometry more accurately by also reducing floaters significantly. With the help of PGSR we achieve view-consistent surfaces and reduce overfitting significantly.}
    \label{tab:evaluation_3dgs_pgsr}
\end{table}
\begin{figure*}[t]
\centering
\setlength{\tabcolsep}{.5pt}
\small
\begin{tabular}{r @{\hspace{0.3cm}} c c c c}
    \multicolumn{1}{l}{\textbf{Scene}} &
    \multicolumn{2}{c}{\textbf{3DGS~\cite{3dgs2023}}} &
    \multicolumn{2}{c}{\textbf{PGSR~\cite{pgsr2024}}}\\
    \multicolumn{1}{c}{\textbf{}} &
    \multicolumn{1}{c}{\textbf{7K Iterations}} &
    \multicolumn{1}{c}{\textbf{15K Iterations}} &
    \multicolumn{1}{c}{\textbf{7K Iterations}} &
    \multicolumn{1}{c}{\textbf{15K Iterations}}\\[0.2cm]\vspace{-0.25em}
    \raisebox{4.0\height}{\textbf{Barn~\cite{tankstemples}}} &
    \includegraphics[width=.19\linewidth]{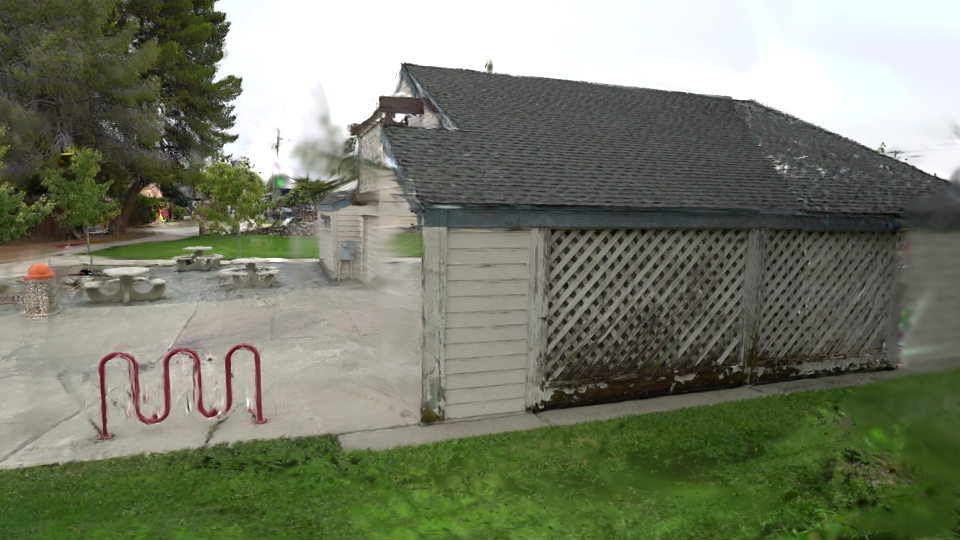} &
    \includegraphics[width=.19\linewidth]{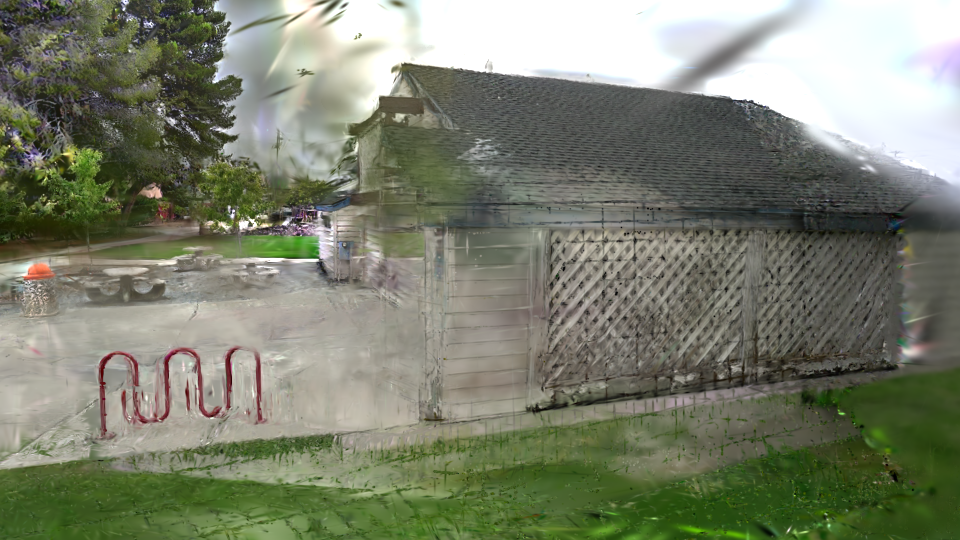} &
    \includegraphics[width=.19\linewidth]{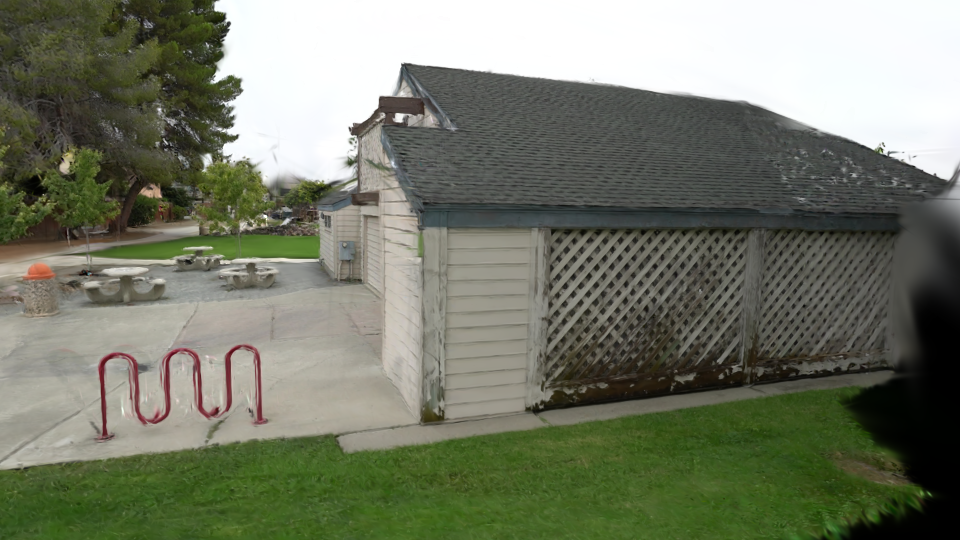} &
    \includegraphics[width=.19\linewidth]{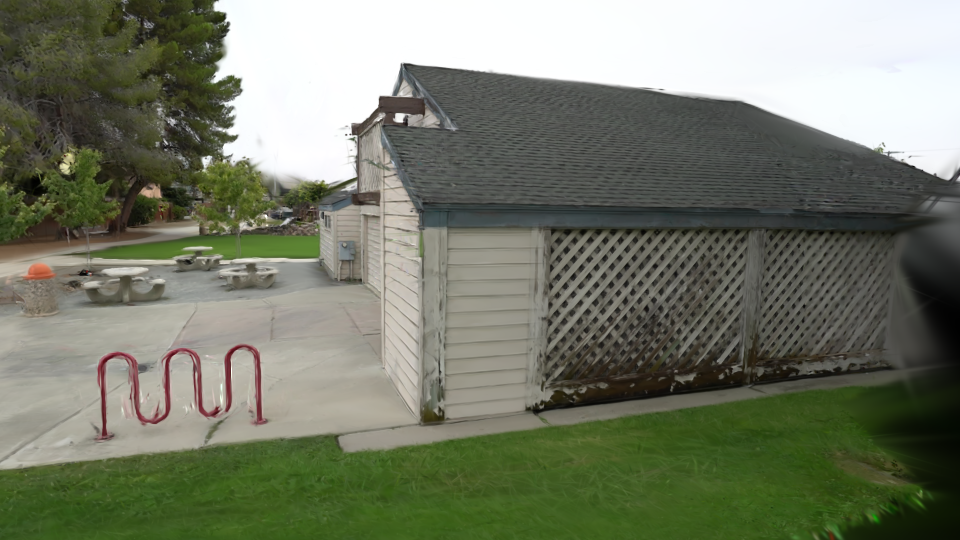}\\\vspace{-0.25em}
    \raisebox{4.0\height}{\textbf{Ballroom~\cite{tankstemples}}} &
    \includegraphics[width=.19\linewidth]{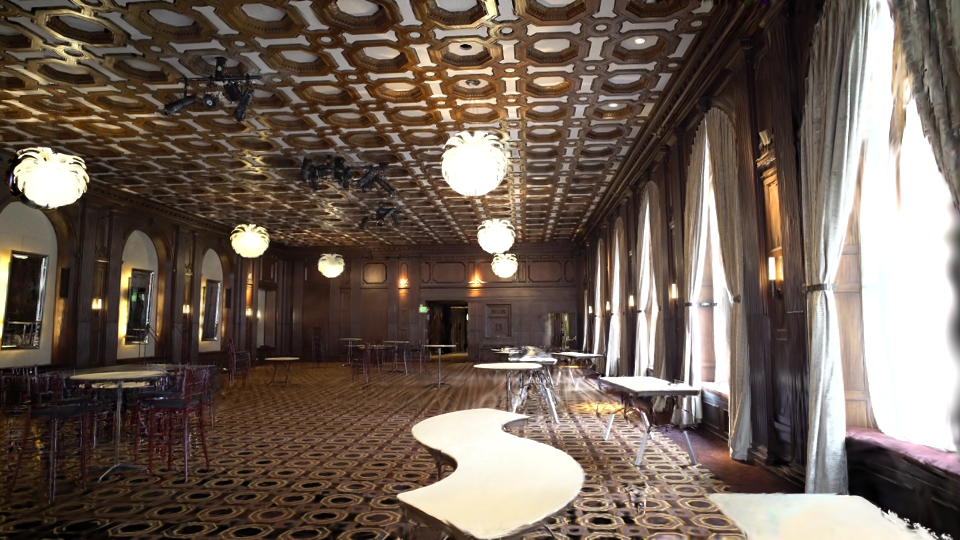} &
    \includegraphics[width=.19\linewidth]{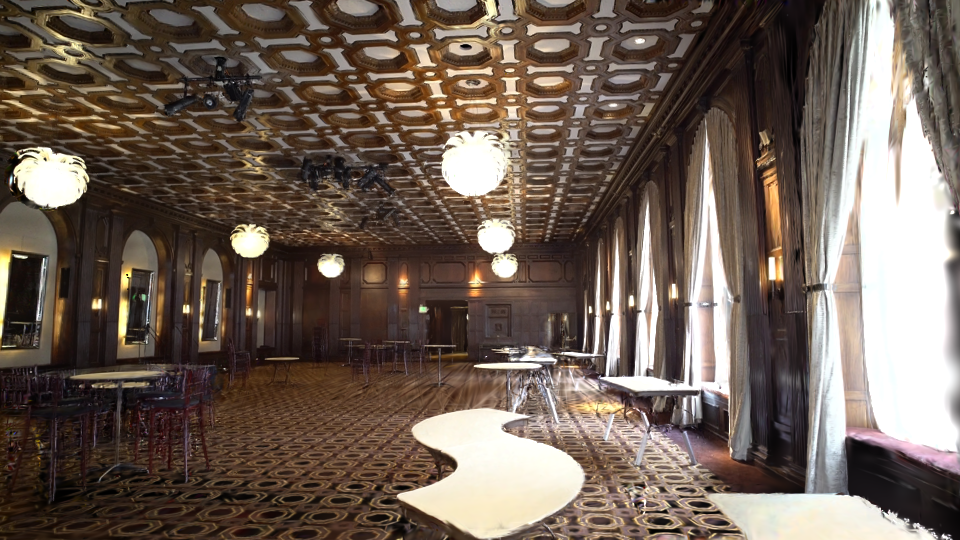} &
    \includegraphics[width=.19\linewidth]{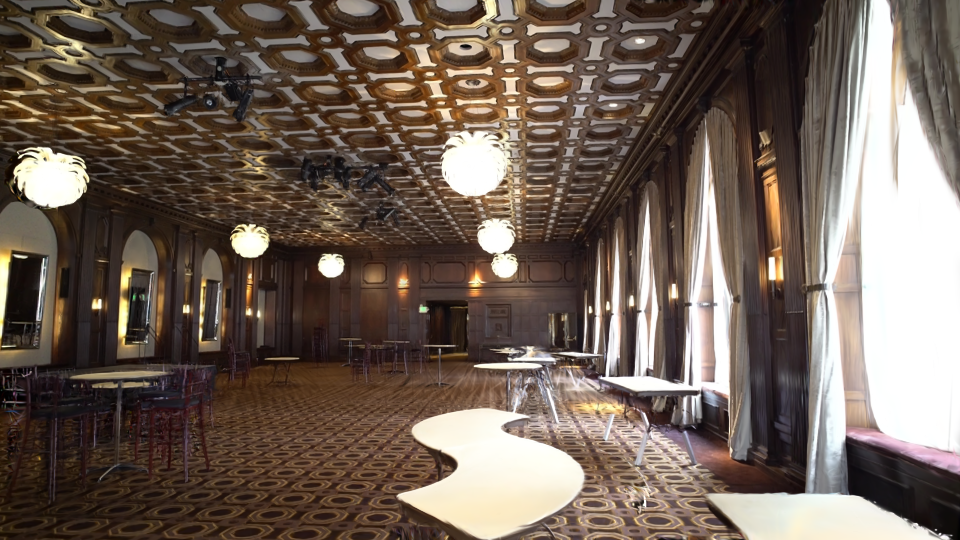} &
    \includegraphics[width=.19\linewidth]{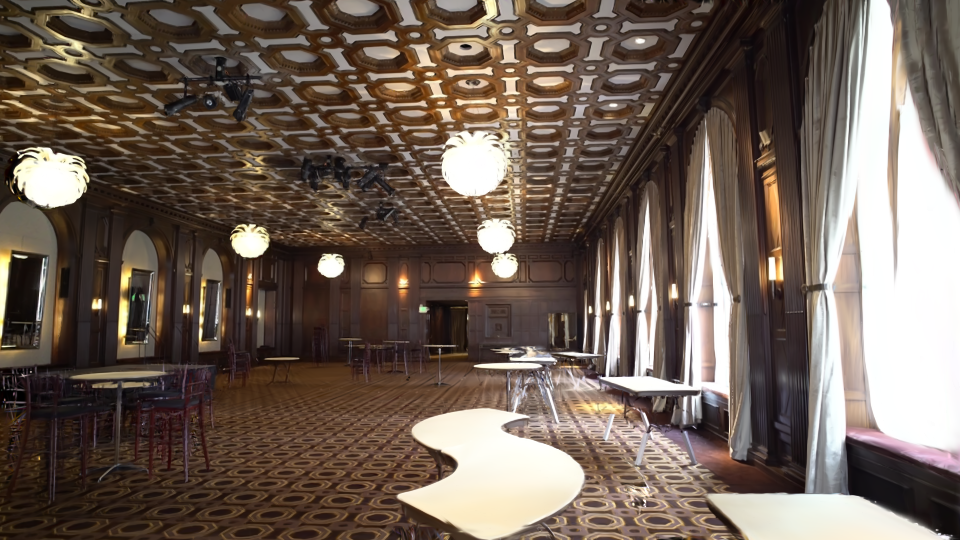}\\\vspace{-0.25em}
    \raisebox{4.0\height}{\textbf{Kitchen~\cite{mipnerf360}}} &
    \includegraphics[width=.19\linewidth]{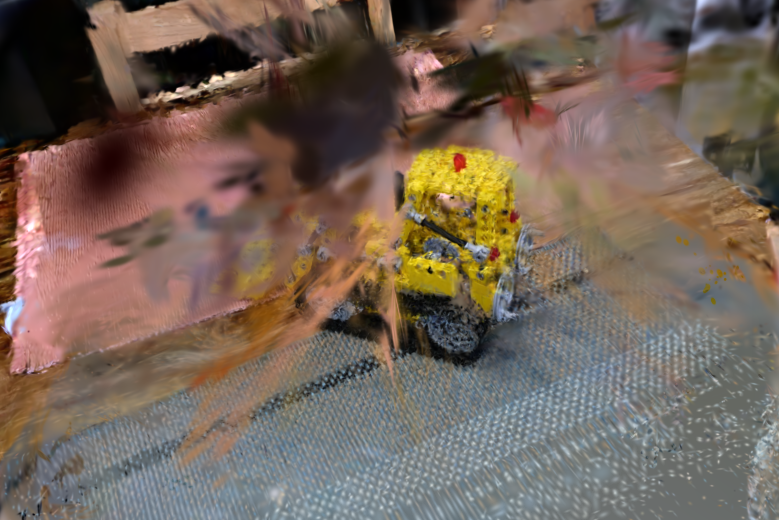} &
    \includegraphics[width=.19\linewidth]{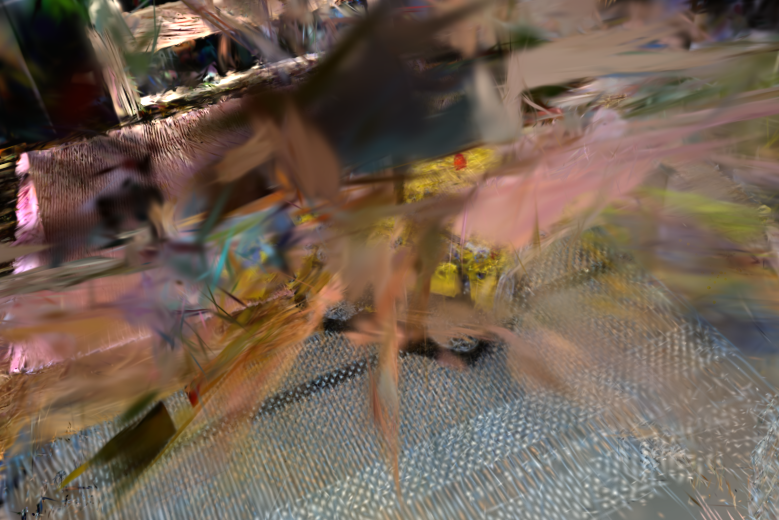} &
    \includegraphics[width=.19\linewidth]{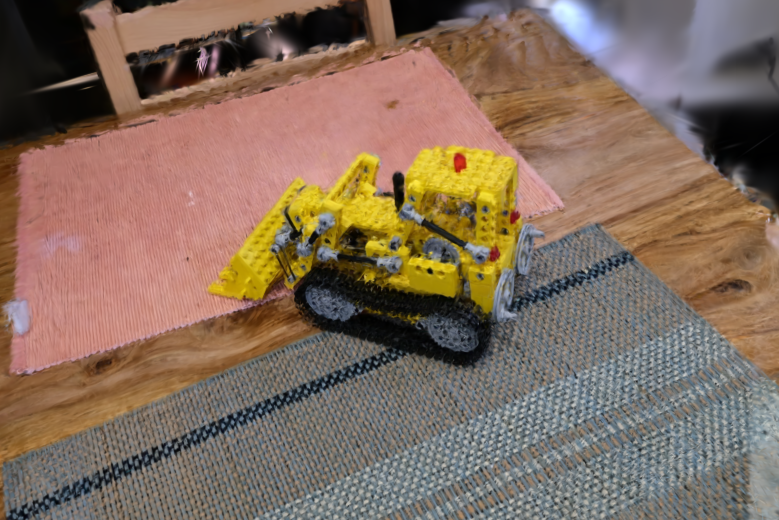} &
    \includegraphics[width=.19\linewidth]{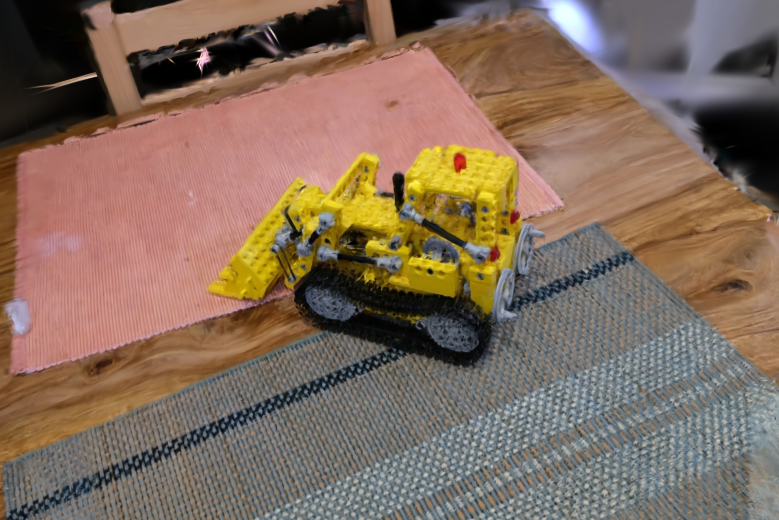}\\\vspace{-0.25em}
    \raisebox{4.0\height}{\textbf{Garden~\cite{mipnerf360}}} &
    \includegraphics[width=.19\linewidth]{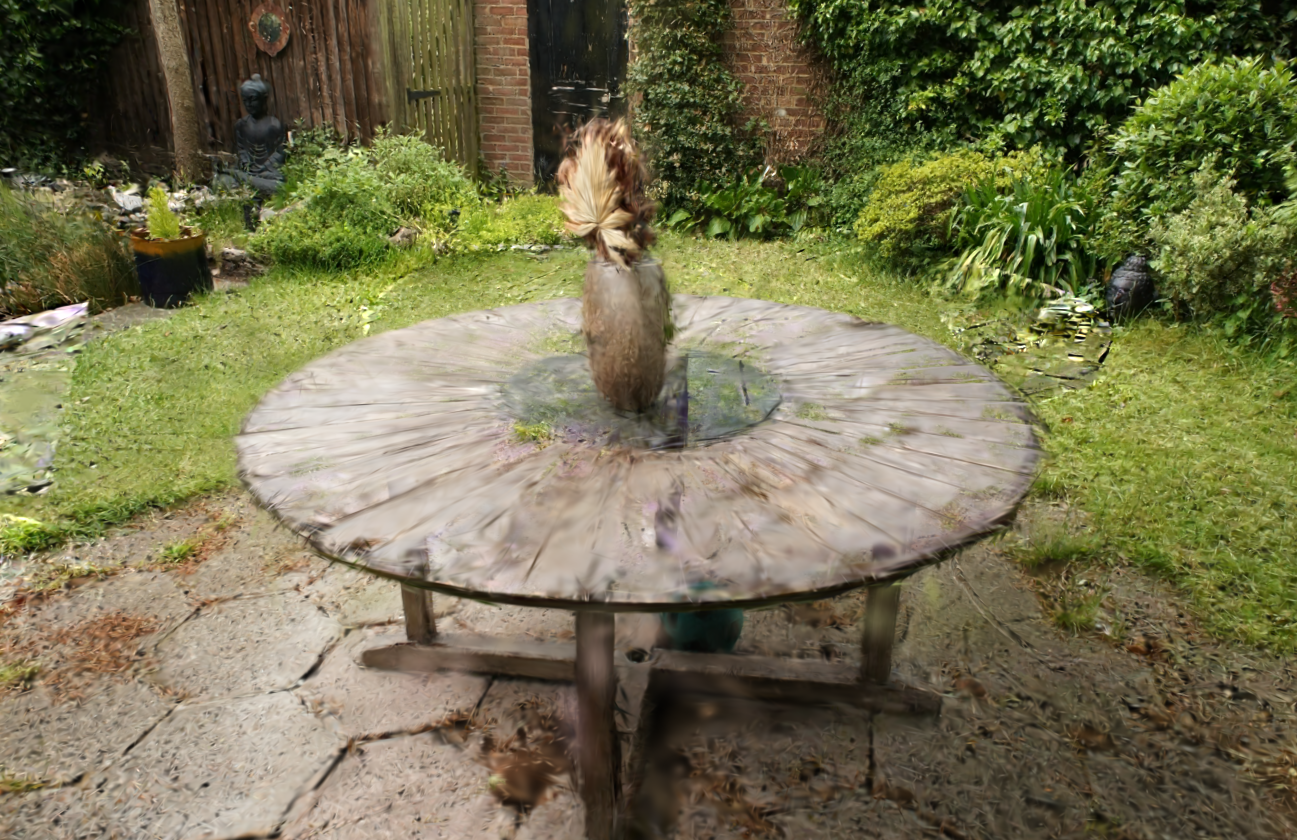} &
    \includegraphics[width=.19\linewidth]{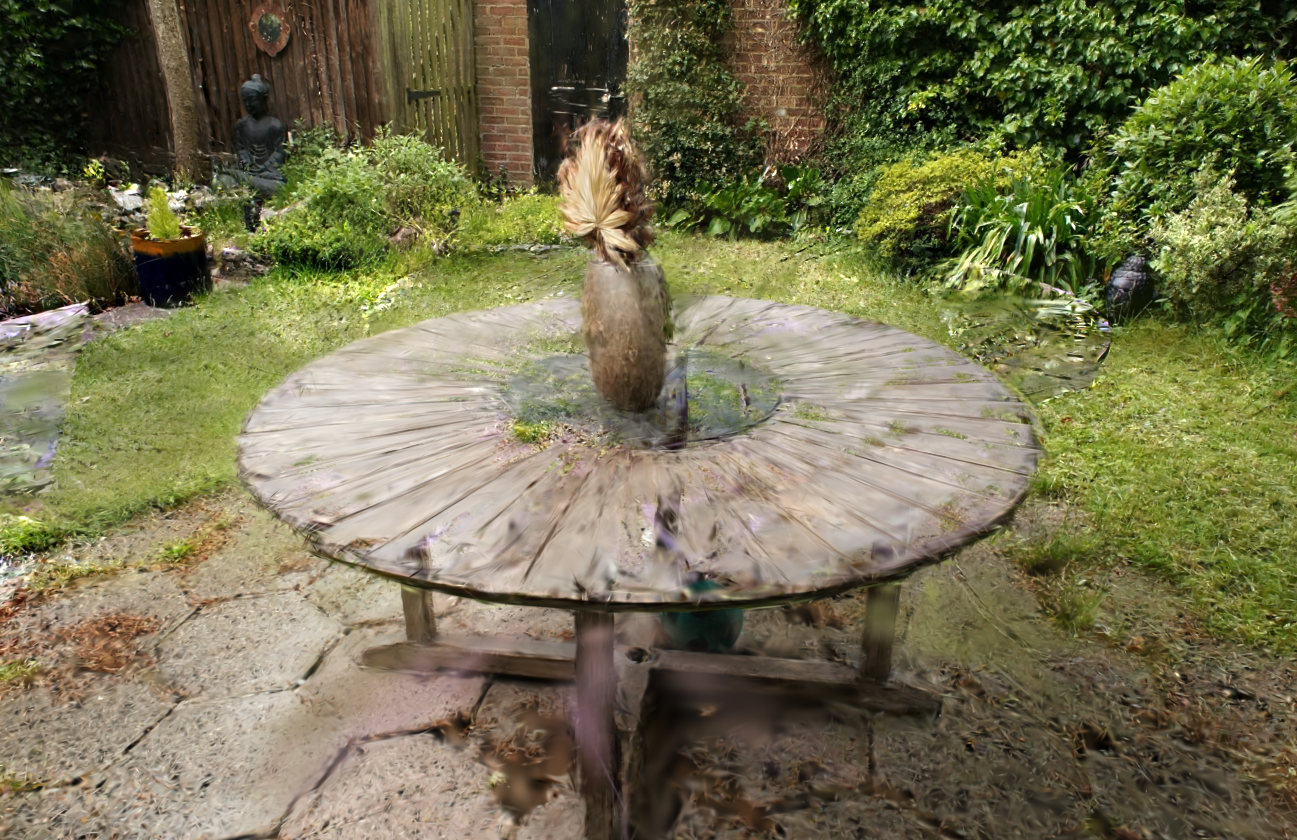} &
    \includegraphics[width=.19\linewidth]{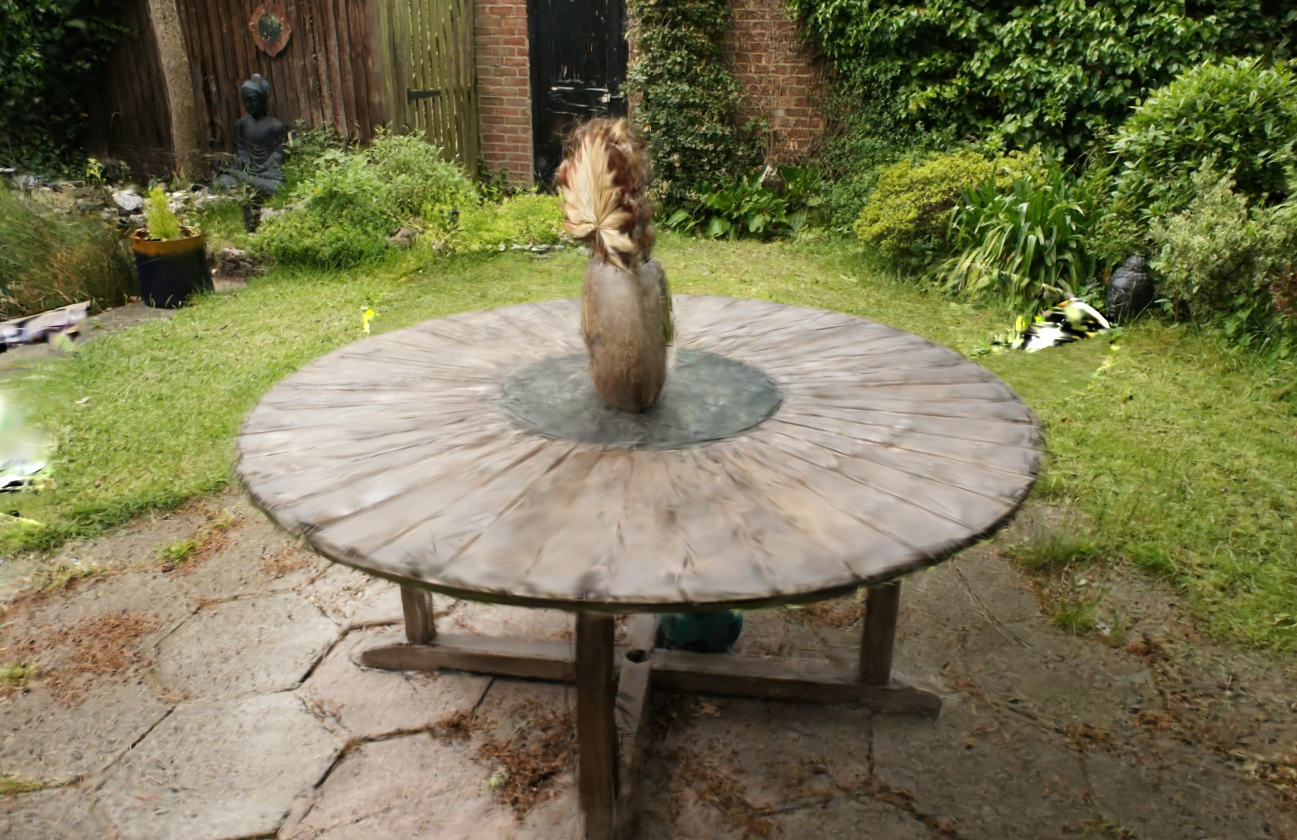} &
    \includegraphics[width=.19\linewidth]{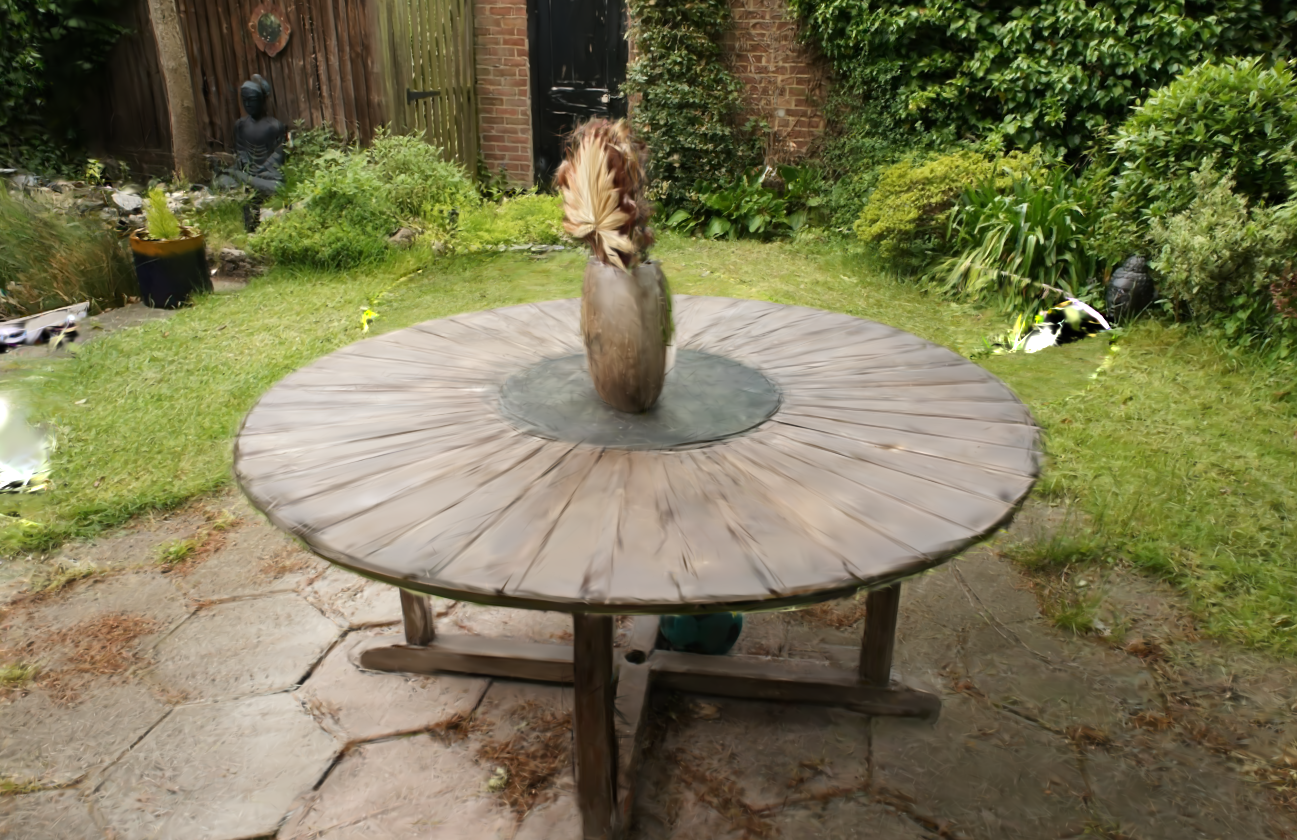}\\\vspace{-0.25em}
\end{tabular}
\vspace{-0.2cm}
\caption{Visual comparison between 3DGS and PGSR with different Iterations. The planar depth from PGSR helps significantly to remove floaters and align the Gaussians accurately to the ground-truth geometry.}
\label{fig:pgsr_vs_3dgs}
\end{figure*}

\begin{table}[ht!]
    \centering
    \footnotesize
    \setlength{\tabcolsep}{2pt}
    \begin{tabular}{lccc}
    \toprule
     Method & PSNR\textsuperscript{$\uparrow$} & SSIM\textsuperscript{$\uparrow$} & LPIPS\textsuperscript{$\downarrow$}\\
    \midrule
 3DGS~\cite{3dgs2023} & 15.36 & 0.572 & 0.379 \\
 DNGaussian~\cite{DNGaussian} & 20.69 & 0.721 & 0.277 \\
 SparseGS~\cite{SparseGS} & 21.20 & 0.717 & 0.231 \\
 InstantSplat~\cite{InstantSplat} & 22.20 & \cellcolor{tab_color!30}{0.743} & 0.199 \\
 FSGS~\cite{fsgs}&\cellcolor{tab_color!10}22.31& 0.693&\cellcolor{tab_color!10}0.197\\
 Intern-GS~\cite{InternGS} & \cellcolor{tab_color!30}{22.67} & \cellcolor{tab_color!10}0.736 & \cellcolor{tab_color!30}{0.191} \\
 Ours & \cellcolor{tab_color!50}{22.87} & \cellcolor{tab_color!50}{0.764} & \cellcolor{tab_color!50}{0.189}\\
    \bottomrule
    \end{tabular}
    \caption{Evaluation on Tanks and Temples dataset with 3-view setting. Our model does not oversmooth high-frequency textures and accurately aligns the Gaussians with the underlying surface geometry.}
    \label{tab:evaluation_tat}
\end{table}
\begin{table}[ht!]
    \centering
    \footnotesize
    \setlength{\tabcolsep}{2pt}
    \begin{tabular}{lccc}
    \toprule
    Method & PSNR\textsuperscript{$\uparrow$} & SSIM\textsuperscript{$\uparrow$} & LPIPS\textsuperscript{$\downarrow$}\\
    \midrule
 MASt3R Initialization~\cite{GaussianScenes} & 12.59 & 0.231 & 0.593 \\ 
 Gaussian Scenes~\cite{GaussianScenes} & \cellcolor{tab_color!10}{13.81} & \cellcolor{tab_color!10}{0.265} & \cellcolor{tab_color!30}{0.547} \\ 
 FSGS~\cite{fsgs} & \cellcolor{tab_color!50}14.17 & \cellcolor{tab_color!50}0.318 & \cellcolor{tab_color!10}0.578 \\
 Ours & \cellcolor{tab_color!30}14.14 & \cellcolor{tab_color!30}{0.310} & \cellcolor{tab_color!50}{0.523}\\
    \bottomrule
    \end{tabular}
    \caption{Evaluation on MipNeRF360 dataset with 3-view setting. Our model reconstructs seen regions accurately, but can not introduce geometry in unseen regions.}
    \label{tab:evaluation_mipnerf360_3view}
\end{table}
\begin{table}[ht!]
    \centering
    \footnotesize
    \setlength{\tabcolsep}{2pt}
    \begin{tabular}{lccc}
    \toprule
     Method & PSNR\textsuperscript{$\uparrow$} & SSIM\textsuperscript{$\uparrow$} & LPIPS\textsuperscript{$\downarrow$}\\
     \midrule
 3DGS~\cite{3dgs2023} & \cellcolor{tab_color!10}17.49 & \cellcolor{tab_color!10}0.490 & \cellcolor{tab_color!10}0.431 \\
 DNGaussian~\cite{DNGaussian} & 16.28 & 0.432 & 0.549 \\
 SparseGS~\cite{SparseGS} & \cellcolor{tab_color!30}{19.37} & \cellcolor{tab_color!50}{0.577} & \cellcolor{tab_color!30}{0.398} \\
 Ours & \cellcolor{tab_color!50}{19.54} & \cellcolor{tab_color!30}{0.492} & \cellcolor{tab_color!50}{0.362}\\
    \bottomrule
    \end{tabular}
    \caption{Evaluation on MipNeRF360 dataset with 12-view setting. Our model can reconstruct the scenes with highly accurate surface alignment. The ground is view-consistent, and fewer floating artifacts compared to SparseGS~\cite{SparseGS}.}
    \label{tab:evaluation_mipnerf360_12view}
\end{table}
\begin{table}[ht!]
    \centering
    \footnotesize
    \setlength{\tabcolsep}{2pt}
    \begin{tabular}{lcccccc}
    \toprule
    \multirow{3}{*}{Method} & \multicolumn{3}{c}{LLFF} & \multicolumn{3}{c}{DTU}\\
    \cmidrule(lr){2-4} \cmidrule(lr){5-7}
    & PSNR\textsuperscript{$\uparrow$} & SSIM\textsuperscript{$\uparrow$} & LPIPS\textsuperscript{$\downarrow$}& PSNR\textsuperscript{$\uparrow$} & SSIM\textsuperscript{$\uparrow$} & LPIPS\textsuperscript{$\downarrow$}\\
    \midrule
 3DGS~\cite{3dgs2023} & 15.52 & 0.408 & 0.405&10.99 & 0.585 & 0.313 \\ 
 DNGaussian~\cite{DNGaussian} & 19.12 & 0.591 & \cellcolor{tab_color!10}0.294&\cellcolor{tab_color!10}18.91 & 0.790 & \cellcolor{tab_color!10}0.176 \\ 
 SparseGS~\cite{SparseGS} & 19.86 & \cellcolor{tab_color!30}{0.668} & 0.322& 18.89 & \cellcolor{tab_color!30}0.834 & 0.178 \\ 
 InstantSplat~\cite{InstantSplat} & 17.67 & 0.603 & 0.379&17.55 & 0.634 & 0.212 \\ 
 FSGS~\cite{fsgs}&\cellcolor{tab_color!30}20.31& 0.652&0.288&19.54& 0.732&0.199\\
 Intern-GS~\cite{InternGS} & \cellcolor{tab_color!50}{20.49} &  \cellcolor{tab_color!50}{0.693} & \cellcolor{tab_color!50}{0.212}&\cellcolor{tab_color!30}{20.34} & \cellcolor{tab_color!50}{0.851} & \cellcolor{tab_color!30}{0.163} \\ 
 Ours & \cellcolor{tab_color!10}19.92 & \cellcolor{tab_color!10}0.664 & \cellcolor{tab_color!30}{0.254}&\cellcolor{tab_color!50}{23.52} & \cellcolor{tab_color!10}{0.815} & \cellcolor{tab_color!50}{0.145}\\
    \bottomrule
    \end{tabular}
    \caption{Evaluation on LLFF and DTU dataset with 3-view setting. Following previous work, for evaluation on DTU the background masks are applied. Our model is able to reconstruct fine-grained textures accurately, but it underperforms in unobserved regions compared to methods that generate content for unseen regions.}
    \label{tab:llff_dtu_comparison}
\end{table}

\section{Conclusion and Limitations}
Our model shows strong performance under sparse-view constraints, specifically when handling between 3 and 12 views.
The model demonstrates the importance of accurate dense point cloud initialization.
We introduce a modified depth loss that enables correct scene generalization by reducing depth ambiguities without introducing artifacts in low confidence regions.
In addition, we introduce normal and depth warping loss terms that improve alignment with the ground-truth surface geometry. Finally, we relax certain assumptions from PGSR to allow robust optimization in sparse-view settings.

Our model faces limitations when dealing with large datasets, as processing many input views with $\pi^3$ consumes a large amount of GPU memory, which is infeasible on consumer hardware. Additional limitations come from inaccurate depth estimations in specific scenes, such as the leaves scene from the LLFF dataset~\cite{mildenhall2019llff}. Future improvements could include the joint optimization of the camera poses and the Gaussian scene, which would result in improved reconstruction quality. Furthermore, the integration of generative priors could enhance the model's ability to maintain photometric and geometric consistency across occluded or sparse areas.
\FloatBarrier

{
    \small
    \bibliographystyle{ieeenat_fullname}
    \bibliography{references}

@String(CVPR  = {Proc. CVPR})

@String(ICCV  = {Proc. ICCV})

@String(ECCV  = {Proc. ECCV})

@String(CVPRW= {Proc. CVPRW})

@String(P3DV= {Proc. 3DV})

@string{SGP = {Proc. SGP}}

@string{DOCS = {Proc. DOCS}}

@string{EAAI = {Engineering Applications of Artificial Intelligence}}

@string{ECCVW = {ECCV 2024 Workshops}}

@String(IJCV  = {IJCV})

@String(TOG   = {ACM TOG})

@string{ACMC = {Communications of the ACM}}

@String(TVCG  = {IEEE TVCG})

@string{RS = {Remote Sensing}}

@string{AN = {Acta Numerica}}

@article{pgsr2024,
    author = {Chen, Danpeng and Li, Hai and Ye, Weicai and Wang, Yifan and Xie, Weijian and Zhai, Shangjin and Wang, Nan and Liu, Haomin and Bao, Hujun and Zhang, Guofeng},
    title = {{PGSR: Planar-Based Gaussian Splatting for Efficient and High-Fidelity Surface Reconstruction}},
    journal = TVCG,
    year = {2024}
}

@article{3dgs2023,
    author = {Kerbl, Bernhard and Kopanas, Georgios and Leimkuehler, Thomas and Drettakis, George},
    title = {{3D Gaussian Splatting for Real-Time Radiance Field Rendering}},
    journal = TOG,
    year = 2023
}

@inproceedings{DNGaussian,
    author    = {Li, Jiahe and Zhang, Jiawei and Bai, Xiao and Zheng, Jin and Ning, Xin and Zhou, Jun and Gu, Lin},
    title     = {{DNGaussian: Optimizing Sparse-View 3D Gaussian Radiance Fields with Global-Local Depth Normalization}},
    booktitle = CVPR,
    month     = jun,
    year      = {2024},
    pages     = {20775-20785}
}

@inproceedings{SparseGS,
  author={Xiong, Haolin and Muttukuru, Sairisheek and Xiao, Hanyuan and Upadhyay, Rishi and Chari, Pradyumna and Zhao, Yajie and Kadambi, Achuta},
  booktitle=P3DV, 
  title={{SparseGS: Sparse View Synthesis Using 3D Gaussian Splatting}}, 
  year={2025},
  volume={},
  number={},
  pages={1032-1041},
}

@article{InstantSplat,
    title={{InstantSplat: Sparse-view Gaussian Splatting in Seconds}}, 
    author={Zhiwen, Fan and Kairun, Wen and Wenyan, Cong and Kevin, Wang and Jian, Zhang and Xinghao, Ding and Danfei, Xu and Boris, Ivanovic and Marco, Pavone and Georgios, Pavlakos and Zhangyang, Wang and Yue, Wang},
    year={2024},
    journal={arXiv preprint arXiv:2403.20309},
    eprint={arXiv:2403.20309},
    archivePrefix={arXiv},
    primaryClass={cs.CV},
    url={https://arxiv.org/abs/2403.20309}
}

@article{InternGS,
    title={{Intern-GS: Vision Model Guided Sparse-View 3D Gaussian Splatting}}, 
    author={Xiangyu, Sun and Runnan, Chen and Mingming, Gong and Dong, Xu and Tongliang, Liu},
    year={2025},
    journal={arXiv preprint arXiv:2505.20729},
    eprint={arXiv:2505.20729},
    archivePrefix={arXiv},
    primaryClass={cs.CV},
    url={https://arxiv.org/abs/2505.20729}
}

@article{GaussianScenes,
    title={{Gaussian Scenes: Pose-Free Sparse-View Scene Reconstruction using Depth-Enhanced Diffusion Priors}}, 
    author={Paul, Soumava and Kaushik, Prakhar and Yuille, Alan},
    year={2024},
    journal={arXiv preprint arXiv:2411.15966},
    eprint={arXiv:2411.15966},
    archivePrefix={arXiv},
    primaryClass={cs.CV},
    url={https://arxiv.org/abs/2411.15966}
}

@article{wang2025pi3,
      title={{$\pi^3$: Scalable Permutation-Equivariant Visual Geometry Learning}}, 
      author={Yifan Wang and Jianjun Zhou and Haoyi Zhu and Wenzheng Chang and Yang Zhou and Zizun Li and Junyi Chen and Jiangmiao Pang and Chunhua Shen and Tong He},
      year={2025},
    journal={arXiv preprint arXiv:2507.13347},
      eprint={arXiv:2507.13347},
      archivePrefix={arXiv},
      primaryClass={cs.CV},
      url={https://arxiv.org/abs/2507.13347}
}

@article{tankstemples,
    author = {Knapitsch, Arno and Park, Jaesik and Zhou, Qian-Yi and Koltun, Vladlen},
    title = {{Tanks and Temples: Benchmarking Large-Scale Scene Reconstruction}},
    year = {2017},
    issue_date = {August 2017},
    publisher = {Association for Computing Machinery},
    address = {New York, NY, USA},
    volume = {36},
    number = {4},
    issn = {0730-0301},
    url = {https://doi.org/10.1145/3072959.3073599},
    doi = {10.1145/3072959.3073599},
    journal = TOG,
    month = jul,
    articleno = {78},
    numpages = {13}
}

@article{mildenhall2019llff,
  title={{Local Light Field Fusion: Practical View Synthesis with Prescriptive Sampling Guidelines}},
  author={Ben Mildenhall and Pratul P. Srinivasan and Rodrigo Ortiz-Cayon and Nima Khademi Kalantari and Ravi Ramamoorthi and Ren Ng and Abhishek Kar},
  journal=TOG,
  year={2019},
}

@article{dtu,
  title={{Large-Scale Data for Multiple-View Stereopsis}},
  author={Aan{\ae}s, Henrik and Jensen, Rasmus Ramsb{\o}l and Vogiatzis, George and Tola, Engin and Dahl, Anders Bjorholm},
  journal=IJCV,
  pages={1--16},
  year={2016},
  publisher={Springer}
}

@inproceedings{mipnerf360,
    author    = {Barron, Jonathan T. and Mildenhall, Ben and Verbin, Dor and Srinivasan, Pratul P. and Hedman, Peter},
    title     = {{Mip-NeRF 360: Unbounded Anti-Aliased Neural Radiance Fields}},
    booktitle = CVPR,
    month     = jun,
    year      = {2022},
    pages     = {5470-5479}
}

@inproceedings{EVALScenReconstruction2024,
  author={Zhou, Yiming and Zeng, Zixuan and Chen, Andi and Zhou, Xiaofan and Ni, Haowei and Zhang, Shiyao and Li, Panfeng and Liu, Liangxi and Zheng, Mengyao and Chen, Xupeng},
  booktitle=DOCS, 
  title={{Evaluating Modern Approaches in 3D Scene Reconstruction: NeRF vs Gaussian-Based Methods}}, 
  year={2024},
  volume={},
  number={},
  pages={926-931}
}

@article{SLAMROBOT2021,
title = {{SLAM; definition and evolution}},
journal = EAAI,
volume = {97},
pages = {104032},
year = {2021},
issn = {0952-1976},
doi = {https://doi.org/10.1016/j.engappai.2020.104032},
url = {https://www.sciencedirect.com/science/article/pii/S0952197620303092},
author = {Hamid Taheri and Zhao Chun Xia}
}

@article{NeRFOriginal,
author = {Mildenhall, Ben and Srinivasan, Pratul P. and Tancik, Matthew and Barron, Jonathan T. and Ramamoorthi, Ravi and Ng, Ren},
title = {{NeRF: Representing Scenes as Neural Radiance Fields for View Synthesis}},
year = {2021},
issue_date = {January 2022},
publisher = {Association for Computing Machinery},
address = {New York, NY, USA},
volume = {65},
number = {1},
issn = {0001-0782},
url = {https://doi.org/10.1145/3503250},
doi = {10.1145/3503250},
journal = ACMC,
month = dec,
pages = {99–106},
numpages = {8}
}

@inproceedings{few-shotNVS3DGS,
author={Kumar, Raja
and Vats, Vanshika},
editor={Del Bue, Alessio
and Canton, Cristian
and Pont-Tuset, Jordi
and Tommasi, Tatiana},
title={{Few-Shot Novel View Synthesis Using Depth Aware 3D Gaussian Splatting}},
booktitle=ECCVW,
year={2025},
publisher={Springer Nature Switzerland},
address={Cham},
pages={1--13},
isbn={978-3-031-91989-3}
}

@inproceedings{
    mastrsfm2025,
    title={{{MAS}t3R-SfM: a Fully-Integrated Solution for Unconstrained Structure-from-Motion}},
    author={Bardienus Pieter Duisterhof and Lojze Zust and Philippe Weinzaepfel and Vincent Leroy and Yohann Cabon and Jerome Revaud},
    booktitle=P3DV,
    year={2025},
    url={https://openreview.net/forum?id=5uw1GRBFoT}
}

@inproceedings{dust3r_cvpr24,
      title={{DUSt3R: Geometric 3D Vision Made Easy}},
      author={Shuzhe Wang and Vincent Leroy and Yohann Cabon and Boris Chidlovskii and Jerome Revaud},
      booktitle = CVPR,
      year = {2024}
}

@article{dssim,
  author={Baker, Allison H. and Pinard, Alexander and Hammerling, Dorit M.},
  journal=TVCG, 
  title={{On a Structural Similarity Index Approach for Floating-Point Data}}, 
  year={2024},
  volume={30},
  number={9},
  pages={6261-6274},
  doi={10.1109/TVCG.2023.3332843}
}

@inproceedings{
    gaussianproDepth,
    title={{GaussianPro: 3D Gaussian Splatting with Progressive Propagation}},
    author={Kai Cheng and Xiaoxiao Long and Kaizhi Yang and Yao Yao and Wei Yin and Yuexin Ma and Wenping Wang and Xuejin Chen},
    booktitle={Proceedings of the 41st International Conference on Machine Learning (ICML 2024)},
    year={2024},
    url={https://openreview.net/forum?id=lQ3SEBH1gF}
}

@inproceedings{SFM-Original,
    author={Sch\"{o}nberger, Johannes Lutz and Frahm, Jan-Michael},
    title={{Structure-from-Motion Revisited}},
    booktitle=CVPR,
    year={2016},
}

@inproceedings{MVSPixel,
    author={Sch\"{o}nberger, Johannes Lutz and Zheng, Enliang and Pollefeys, Marc and Frahm, Jan-Michael},
    title={{Pixelwise View Selection for Unstructured Multi-View Stereo}},
    booktitle=ECCV,
    year={2016},
}

@article{NeRF-Analysis,
AUTHOR = {Remondino, Fabio and Karami, Ali and Yan, Ziyang and Mazzacca, Gabriele and Rigon, Simone and Qin, Rongjun},
TITLE = {{A Critical Analysis of NeRF-Based 3D Reconstruction}},
JOURNAL = RS,
VOLUME = {15},
YEAR = {2023},
NUMBER = {14},
ARTICLE-NUMBER = {3585},
URL = {https://www.mdpi.com/2072-4292/15/14/3585},
ISSN = {2072-4292},
DOI = {10.3390/rs15143585}
}

@article{SFM-Survey,
title={{A Survey of Structure from Motion}},
volume={26},
DOI={10.1017/S096249291700006X},
journal=AN,
author={Özyeşil, Onur and Voroninski, Vladislav and Basri, Ronen and Singer, Amit},
year={2017}, pages={305–364}
}

@inproceedings{kazhdan2006poisson,
  title={{Poisson surface reconstruction}},
  author={Kazhdan, Michael and Bolitho, Matthew and Hoppe, Hugues},
  booktitle=SGP,
  year={2006}
}

@article{mueller2022instantngp,
    author = {Thomas M\"uller and Alex Evans and Christoph Schied and Alexander Keller},
    title = {{Instant Neural Graphics Primitives with a Multiresolution Hash Encoding}},
    journal = TOG,
    issue_date = {July 2022},
    volume = {41},
    number = {4},
    month = jul,
    year = {2022},
    pages = {102:1--102:15},
    articleno = {102},
    numpages = {15},
    url = {https://doi.org/10.1145/3528223.3530127},
    doi = {10.1145/3528223.3530127},
    publisher = {ACM},
    address = {New York, NY, USA},
}

@inproceedings{PlenOctree,
    author    = {Yu, Alex and Li, Ruilong and Tancik, Matthew and Li, Hao and Ng, Ren and Kanazawa, Angjoo},
    title     = {{PlenOctrees for Real-Time Rendering of Neural Radiance Fields}},
    booktitle = ICCV,
    month     = oct,
    year      = {2021},
    pages     = {5752-5761}
}

@inproceedings{EfficientNerf,
    author    = {Hu, Tao and Liu, Shu and Chen, Yilun and Shen, Tiancheng and Jia, Jiaya},
    title     = {{EfficientNeRF: Efficient Neural Radiance Fields}},
    booktitle = CVPR,
    month     = jun,
    year      = {2022},
    pages     = {12902-12911}
}

@inproceedings{depthreg3dgs,
    author    = {Chung, Jaeyoung and Oh, Jeongtaek and Lee, Kyoung Mu},
    title     = {{Depth-Regularized Optimization for 3D Gaussian Splatting in Few-Shot Images}},
    booktitle = CVPRW,
    month     = jun,
    year      = {2024},
    pages     = {811-820}
}

@inproceedings{DropGaussian,
    author    = {Park, Hyunwoo and Ryu, Gun and Kim, Wonjun},
    title     = {{DropGaussian: Structural Regularization for Sparse-view Gaussian Splatting}},
    booktitle = CVPR,
    month     = jun,
    year      = {2025},
    pages     = {21600-21609}
}

@inproceedings{DropoutGS,
    author    = {Xu, Yexing and Wang, Longguang and Chen, Minglin and Ao, Sheng and Li, Li and Guo, Yulan},
    title     = {{DropoutGS: Dropping Out Gaussians for Better Sparse-view Rendering}},
    booktitle = CVPR,
    month     = jun,
    year      = {2025},
    pages     = {701-710}
}

@inproceedings{COLMAPFREE3DGS,
    author    = {Fu, Yang and Liu, Sifei and Kulkarni, Amey and Kautz, Jan and Efros, Alexei A. and Wang, Xiaolong},
    title     = {{COLMAP-Free 3D Gaussian Splatting}},
    booktitle = CVPR,
    month     = jun,
    year      = {2024},
    pages     = {20796-20805}
}

@inproceedings{fsgs,
author = {Zhu, Zehao and Fan, Zhiwen and Jiang, Yifan and Wang, Zhangyang},
title = {FSGS: Real-Time Few-Shot View Synthesis Using Gaussian Splatting},
year = {2024},
isbn = {978-3-031-72932-4},
publisher = {Springer-Verlag},
address = {Berlin, Heidelberg},
url = {https://doi.org/10.1007/978-3-031-72933-1_9},
doi = {10.1007/978-3-031-72933-1_9},
booktitle = ECCV,
pages = {145–163},
numpages = {19},
location = {Milan, Italy}
}

@InProceedings{GenFusion,
    author    = {Wu, Sibo and Xu, Congrong and Huang, Binbin and Geiger, Andreas and Chen, Anpei},
    title     = {GenFusion: Closing the Loop between Reconstruction and Generation via Videos},
    booktitle = CVPR,
    month     = {June},
    year      = {2025},
    pages     = {6078-6088}
}
}

\end{document}